\documentclass[ twocolumn,aps,prd,   
               preprintnumbers,numbers,sort&compress,
               nofootinbib,
                            showpacs,
               colorlinks,
               linkcolor=blue,
               citecolor=blue]{revtex4-1}
   \newcommand{\exclude}[1]{}

\usepackage{graphicx,amsmath,amssymb,bm}
\usepackage{psfrag}
 \usepackage{feynmp}
\usepackage{hyperref}
\usepackage{enumitem}

\newcommand{\beq}{\begin{equation}}
\newcommand{\eeq}{\end{equation}}
\newcommand{\be}{\begin{eqnarray}}
\newcommand{\ee}{\end{eqnarray}}
   
\def\dd{ \,\mathrm{d} }

\def\+{\dagger}
 \def\la{\langle}
 \def\ra{\rangle}

\begin{document}

\title{ Maxwell Theory on a Compact Manifold as a Topologically Ordered System }

\author{  Ariel R. Zhitnitsky}  

\affiliation{ Department of Physics \& Astronomy, University of British Columbia, Vancouver, B.C. V6T 1Z1, Canada} 

\begin{abstract}
  We study  novel type of contributions to  the   entropy of  the Maxwell system defined on a compact manifold such as torus. These new terms   are not related to the physical propagating photons.  Rather, these novel contributions  emerge as a result of   tunnelling events  when transitions occur between topologically different but physical identical vacuum winding states. We compute two  new (topologically protected) types of contributions to  the entropy in this system   resulting from this dynamics. First  contribution  has a negative sign, expressed in terms of the magnetic susceptibility, and it  is similar in spirit to topological entanglement entropy discussed in condensed matter systems. The second contribution 
  with a positive sign results from the emergent degeneracy which occurs when the system   is placed into a background of external magnetic field. This degeneracy resembles a similar effect which occurs at $\theta=\pi$ in topological insulators.     Based on these computations we  claim that the Maxwell system defined  on 4 torus behaves in many respects as a  topologically ordered system.

\pacs{11.15.-q, 11.15.Kc, 11.15.Tk}
 
\end{abstract} 

\maketitle

\section{Introduction. Motivation.}\label{introduction}
 The main motivation for present studies is as follows.
 It has been recently argued \cite{Cao:2013na} that if  the Maxwell theory  is   defined on a small compact manifold  than  some novel  terms in the partition function will emerge.  These terms are not related to the  propagating photons with two transverse physical polarizations. Rather, these novel terms occur as a result  of  tunnelling  events between topologically different but physically identical    states. These states play no role when the system is defined in Minkowski space-time ${\mathbb{R}_{1,3}}$. But these states become important when   the system   is defined on a finite compact manifold such as  torus ${\mathbb{T}}^4$. 
 
 In particular,  it has been explicitly shown in  \cite{Cao:2013na} that these novel terms lead to a fundamentally new  contributions to the Casimir vacuum pressure, which can  not be expressed  in terms of conventional  propagating  physical degrees of freedom.     Instead,  the new contributions   appear  as a result  of tunnelling events between different topological sectors $|k\ra$. 
 Mathematically, these sectors  emerge as a result of non-triviality of    the fundamental group $\pi_1[U(1)]\cong \mathbb{Z}$ 
 when the system is defined on a torus.
  
 The crucial for the present studies observation is as follows. While the Maxwell Electrodynamics is the theory of massless particles (photons), the topological portion of the system decouples from dynamics of these massless propagating photons.  Indeed, as we discuss below, the total partition function ${\cal{Z}}$ can be represented as  a product  ${\cal{Z}} ={\cal{Z}}_{0}\times {\cal{Z}}_{\rm top}$.  The conventional partition function ${\cal{Z}}_{0}$ describing physical photons is  not sensitive  to the topological sectors $|k\ra$ of the system which itself is described by ${\cal{Z}}_{\rm top}$. The topological portion of the partition function ${\cal{Z}}_{\rm top}$
 behaves very much   as topological quantum field theory (TQFT) as we  argue below. Furthermore, it  demonstrates  many features 
 of   topologically ordered systems, which were  initially   introduced in context of condensed matter systems, see  original papers \cite{Wen:1989iv,Wen:1990zza,Moore:1991ks,BF} and recent reviews \cite{Cho:2010rk,Wen:2012hm,Sachdev:2012dq}. 
 
 As a result of these similarities, the key question addressed in the present work is as follows.
 It has been known since \cite{Kitaev:2005dm,Levin:2006zz} that the topologically ordered systems  can be characterized by the so-called topological entanglement   entropy   (TEE).  While the TEE    is a sub-leading contribution to the entanglement entropy, it is nevertheless  a     topologically protected universal constant of the system which can serve as a probe of the topological order. 
 
 We formulate a similar question for the Maxwell system defined on   a non-trivial compact manifold: is there a similar universal contribution to the entropy  
 which is topologically protected and which can serve as a probe of the topological  order? Our ultimate answer is ``yes" as our explicit computations below show.
 Furthermore, this universal constant contribution to the entropy can not be expressed  in terms of physical propagating photons. Instead, it is formulated 
 in terms of the magnetic susceptibility,  which itself is topologically protected object, and which is saturated by the ``instantons", rather than by propagating degrees of freedom.   In many respects this object is  similar to  well known  topological susceptibility in QCD.  As we shall see below this object  does not vanish exclusively  as a result of the dynamics of the topological sectors   described by ${\cal{Z}}_{\rm top}$.
 
 The second question we address in this work can be formulated as follows.  It is known that the  main feature of a topologically ordered system is the presence of a degeneracy of the ground state which can not be   described in terms of  any local observables.    We formulate a similar question for the Maxwell system defined on   a non-trivial compact manifold: is there a similar degeneracy  which can be described by some global,  rather than local, characteristics? Our ultimate answer is ``yes" again, as our explicit computations below show. This degeneracy resembles a similar feature which occurs at $\theta=\pi$ in topological insulators.

One may wonder at this point  what went wrong with the standard and generically accepted arguments 
suggesting  that all physical effects in Maxwell theory  can be formulated in terms of physical propagating photons which are completely 
described by conventional ${\cal{Z}}_{0}$.
Yet, all the effects discussed in the present work are formulated in fundamentally different terms coded by ${\cal{Z}}_{\rm top}$.
The point is that  the standard description is not quite complete  when 4d Maxwell theory is formulated on a non-simply connected, compact manifold. The standard ``naive" argument neglects the topological sectors, which are indeed absent when the theory is formulated  in the  topologically trivial Minkowski space-time. However, these topological sectors    become important when  the  theory is formulated on a nontrivial manifold. 

When one attempts to remove all unphysical degrees of freedom by a  gauge fixing, the 
 physics related to pure gauge configurations describing the  topological sectors of the theory does not go away; instead, this physics  reappears in a much more complicated form where the so-called Gribov's ambiguities \cite{Gribov:1977wm} emerge in Maxwell system formulated on a compact manifold 
 \cite{Killingback:1984en,Parthasarathy:1988sa,Kelnhofer:2007zk,Kelnhofer:2012ig}, see some comments on this matter in concluding section~\ref{conclusion}. In this work we opt to keep some gauge freedom and study these topological sectors explicitly, rather than deal with (technically complicated) analysis of the Gribov's copies.

 The structure of our presentation is as follows. In the next section \ref{topology}, we review the relevant parts of the two dimensional Maxwell ``empty" theory which does not have  any physical propagating degrees of freedom. Still,  it demonstrates a number of  very  nontrivial topological features present in the system.    In section \ref{4d} we generalize our computations for 
 4d Maxwell theory defined  on four torus. We find  two  types of novel contributions to the entropy in this system. First  contribution  with  a negative sign  is very similar  to topological entanglement entropy well-studied in topologically ordered condensed matter systems. The second contribution 
  with a positive sign results from emergent degeneracy which occurs when the system   is placed into a background of external magnetic field which resembles a behaviour of   topological insulators with $\theta=\pi$.  
  
  \section{Maxwell theory in two dimensions as topological QFT}\label{topology}
  The 2d Maxwell model has been solved numerous times using very different techniques, see e.g.\cite{Manton:1985jm,Balachandran:1994vi, SW}. It is known that this is an ``empty" theory in a sense that  it does not support  any propagating degrees of freedom in the bulk of space-time.
  It is also  known that this model can be treated as a conventional topological quantum field theory (TQFT). In particular, this model  can be formulated in terms of     the so-called ``BF"  action involving no metric.   Furthermore, this model   exhibits  many other features such as fractional edge observables which are typical for TQFT, see e.g.\cite{Balachandran:1994vi}. We emphasize on these  properties 
   of the 2d Maxwell theory because the topological portion of the partition function ${\cal{Z}}_{\rm top}$ in our description of 4d Maxwell system, given  in  section \ref{4d},  identically  the same   as the partition function of  2d Maxwell system. As we already mentioned such a relation between the two different systems 
     is a result of  decoupling of physical propagating photons from the topological sectors  in 4d system. 
 
  Our goal here is to review this ``empty" 2d Maxwell theory with nontrivial dynamics of the topological sectors
  when  conventional propagating degrees of freedom are not supported by this system.

  \subsection{Partition function}\label{hamiltonian}
  We consider 2d Maxwell theory defined on the Euclidean torus $S^1\times S^1$ with lengths $L$ and $\beta$ respectively. In the Hamiltonian  framework we choose a $A_0=0$ gauge along with $\partial_1 A_1=0$. This implies that $A_1(t)$ is the only dynamical variable of the system with $E=\dot{A_1}$.

  The spectrum for $\theta$ vacua    is well known \cite{Manton:1985jm}  and it is given by $E_n(\theta)= \frac{1}{2}\left(n+\frac{\theta}{2\pi}\right)^2e^2 L$, such that 
  the corresponding  partition function  takes the form
  \be
 \label{Z_2}
 {\cal{Z}}(V, \theta)=   \sum_{n\in \mathbb{Z}} e^{-\frac{e^2V}{2} \left(n+\frac{\theta}{2\pi}\right)^2},
 \ee
 where $V = \beta L$ is the two-volume of the system.

  We want to reproduce (\ref{Z_2}) using a different approach based on Euclidean path integral computations because it 
  can be easily generalized to similar  computations 4d Maxwell theory defined on 4 torus. Our goal here is to understand the physical meaning of  (\ref{Z_2})  in terms of the path integral computations.

  To proceed with path integral  computations one  considers the  ``instanton"
      configurations   on two dimensional Euclidean torus   with total area $V=L\beta$  described as follows \cite{SW}:
\be
\label{Q}
\int  \dd^2x ~Q(x)  =k,  ~~~~e E^{(k)}=\frac{2\pi k}{V}, 
\ee
 where $Q=\frac{e}{2\pi}E$ is the topological charge density 
and  $k$
is the integer-valued topological charge   in the 2d $U(1)$ gauge theory, $E(x)=\partial_0A_1-\partial_1A_0$ is the field strength. 
The action of this classical configuration is
 \be
 \label{action}
 \frac{1}{2}\int d^2x E^2= \frac{2\pi^2 k^2}{e^2 V}.
 \ee
   This configuration corresponds to the   topological charge $k$ as defined by (\ref{Q}).
The next step is to   compute the  partition function defined as follows
\be
\label{Z_3}
{\cal{Z}}(\theta)=\sum_{ k \in \mathbb{Z}}{\int {\cal{D}}}A^{(k)} {e^{-\frac{1}{2}\int d^2x E^2+ \int d^2x L_{\theta} }},
\ee
where $\theta$ is standard theta parameter which defines the $| \theta\ra$ ground  state and which enters the action with topological density operator
\be
\label{theta}
L_{\theta}=i\theta \int  \dd^2x ~Q(x) =i \theta\frac{e}{2\pi}\int \dd^2x~ E(x). 
\ee
   All integrals in this partition function are gaussian and can be easily evaluated using the technique developed in \cite{SW}. The result is
    \be
 \label{Z_4}
 {\cal{Z}}(V, \theta)= \sqrt{\frac{2\pi}{e^2V}}\sum_{k\in \mathbb{Z}} e^{-\frac{2\pi^2k^2}{e^2V} +ik\theta},
    \ee
    where the expression in the exponent represents the classical instanton configurations with action (\ref{action}) and topological charge (\ref{Q}), while the factor in front is due to the fluctuations, see \cite{Cao:2013na} 
    with some technical details and relevant references. 
      While expressions (\ref{Z_2}) and (\ref{Z_4}) look differently, they are actually identically the same, as the Poisson summation formula states:
     \be
 \label{poisson}
  {\cal{Z}}(\theta)= \sum_{n\in \mathbb{Z}} e^{-\frac{e^2V}{2} \left(n+\frac{\theta}{2\pi}\right)^2}= \sqrt{\frac{2\pi}{e^2V}}\sum_{k\in \mathbb{Z}} e^{-\frac{2\pi^2k^2}{e^2V} +ik\theta}.~~~~
    \ee
   Therefore, we reproduce the original expression (\ref{Z_2})  using the path integral approach. 
   
    The crucial observation for our present study is that this naively ``empty" theory which has no physical propagating degrees of freedom, nevertheless shows some very nontrivial features of the ground state related to the topological properties of the   theory.  These new properties are formulated in terms  of different topological vacuum sectors 
of the system $|k\ra$ which have identical physical properties as they connected to each other by  large gauge transformation operator ${\cal T}$
 commuting   with the Hamiltonian $[{\cal T}, H]=0$. As explained in details in  \cite{Cao:2013na}  the corresponding dynamics of this ``empty" theory represented by partition function (\ref{poisson}) should be interpreted as a result of tunnelling events between  these ``degenerate" winding $ | k\ra$ states which correspond to one and the same physical state.

It is known that this model can be treated as TQFT, e.g. supports edge observables which may assume the  fractional values, and shows many other features which are typical for a TQFT, see \cite{Balachandran:1994vi} and references therein.  The presence of the  topological features of the model can be easily understood from  observation  that entire dynamics of the system is due to the transitions between the topological sectors which themselves are determined by the behaviour of surface integrals at infinity $\oint A_{\mu}dx^{\mu}$. These sectors are classified by   integer numbers 
and they are not sensitive to specific details of the system such as geometrical shape of the system. Therefore, it is not really a surprise that the system is not sensitive to specific geometrical details, and can be treated as TQFT. 

Important  point we would like to make is that our analysis of the topological portion $ {\cal{Z}}_{\rm top}$ of the partition function for   4d Maxwell system defined on ${\mathbb{T}}^4$ assumes exactly the same form (\ref{poisson}) as a result of decoupling of propagating photons from the topological part of the partition function,  as will be discussed in section \ref{4d}. As a result of  this decoupling the topological portion of the 4d Maxwell system behaves very much in the same way as  2d ``empty" theory. Therefore, one should not be very surprised that this 4d system also 
demonstrates some topological features, similar to 2d system reviewed in this section.

Before we proceed with computations of the topological entropy we make a short detour on properties of the 
 topological susceptibility in this model, as it plays an important role in our discussions of the entropy. 
  
  \subsection{Topological susceptibility}\label{top_2}
The topological susceptibility $\chi $  is defined as follows,
  \be
\label{chi1}
\chi \equiv   \lim_{k\rightarrow 0} \int \dd^2x ~e^{ikx}\left< T Q(x) Q(0) \right> ,
\ee
where $Q $ is topological charge density operator normalized according to  eq.(\ref{Q}).  
The   $\chi$  measures response  of the free energy to the introduction of a source  term  
 defined by eq.  (\ref{theta}).
The   computations of $\chi$ in this simple ``empty" model  can be easily carried out  as the partition function ${\cal{Z}}(\theta)$ defined by (\ref{Z_3}) is known exactly (\ref{poisson}). To compute $\chi$ we  should simply differentiate the partition function twice with respect to $\theta$. It leads  to the following well known expression for $\chi$ 
 which is finite  in the infinite volume limit \cite{SW,Zhitnitsky:2011tr}
\be
\label{exact1}
\chi (V\rightarrow \infty)= -\frac{1}{V }\cdot\frac{\partial^2 {\ln\cal{Z}}(\theta)}{\partial\theta^2}|_{\theta=0}= \frac{e^2}{4\pi^2}.  
\ee 
A typical value of the topological charge $k$ which saturates the topological susceptibility $\chi$ in the large volume limit is very large, $k\sim \sqrt{e^2 V}\gg 1$. 
      
   Few comments are in order. First,  any physical state contributes to $\chi $ with negative sign 
\be
\label{dispersion}
\chi_{dispersive} \sim  \lim_{k\rightarrow 0} \sum_n  \frac{\la 0| \frac{e}{2\pi}   E  |n\ra \la n | \frac{e}{2\pi}   E |0\ra }{-k^2-m_n^2} <0, 
\ee
while (\ref{exact1}) has a positive sign. Therefore, 
this non-dispersive (contact) term (\ref{exact1}) can not be identified according to (\ref{dispersion}) with any contribution from any   asymptotic state
even when physical degrees are freedom, such as fermions, are included into the system.
  This term has a fundamentally different, non-dispersive  nature. In fact it is ultimately related to different topological sectors  as our computation (\ref{exact1}) shows. 
  \exclude{
  Secondly, 
 the contact term (\ref{exact1}) can not be removed by any additional prescription of the theory, and it must be present in the system to satisfy the Ward Identity (WI)  which states that $\chi_{QED}(m=0)=0$ when massless fermion is inserted into the system, see \cite{Zhitnitsky:2011tr} for the details.
   Without this   contribution, it would be impossible to satisfy the WI  because the physical propagating degrees of freedom can only contribute with sign $(-)$ to the correlation function as eq. (\ref{dispersion}) suggests, while WI requires $\chi_{QED}(m=0)=0$   in the  chiral limit $m=0$. The WI are automatically satisfied in the system with fermions as a result of exact cancellation between dispersive physical contribution with sign $(-)$ and contact term (\ref{exact1}) with sign $(+)$, see \cite{Zhitnitsky:2011tr}  for the details.
}   
   Secondly, the integrand for the topological susceptibility (\ref{chi1})   demonstrates a singular behaviour 
   \be
   \label{local}
   \left<  Q(x) Q(0) \right>= \frac{e^2}{4\pi^2}\delta^2 (x), 
   \ee
   which is   a not specific property  of this ``empty" theory, but in fact a very generic feature which is present  in many gauge theories; it represents the contact term which is not related to any propagating degrees of freedom.   In particular, such singular behaviour (\ref{local})  is known to  occur in QCD and its modifications, and well supported by the QCD lattice Monte Carlo simulations,   see \cite{Zhitnitsky:2011tr}  for the details and related references. 
   
   The $\delta^2 (x)$ function in  (\ref{local}) should be understood as total divergence related to the infrared (IR) physics, rather than to ultraviolet (UV) behaviour. Indeed,  
\be	\label{divergence}
  \chi  &=& \frac{e^2}{4\pi^2}  \int\delta^2 (x) \dd^2x  
=   \frac{e^2}{4\pi^2}\int   \dd^2x~
  \partial_{\mu}\left(\frac{x^{\mu}}{2\pi x^2}\right)\nonumber\\
  &=&  \frac{e^2}{4\pi^2} \oint_{S_1}    \dd l_{\mu}
 \left(\frac{x^{\mu}}{2\pi x^2}\right).
\ee
  In other words, the non-dispersive contact term with ``wrong" sign (\ref{local}) is  determined by  IR physics
 at arbitrary large distances rather than UV physics which can be erroneously assumed to be  a source of   
 $\delta^2 (x)$ behaviour in (\ref{local}). 
 Our computations in terms of the delocalized instantons (\ref{Q}) explicitly  show that   all observables in this system are originated from the IR physics. 
   
 Our final comment here is that the same contact  term (\ref{exact1}) and its local expression (\ref{local}) can be also computed  using the auxiliary ghost field, the so-called Kogut-Susskind (KS) ghost,  as it has been originally done in ref. \cite{KS}, see also \cite{Zhitnitsky:2011tr} for relevant  discussions in the present context. This auxiliary KS ghost field  provides  the required ``wrong" sign  (\ref{exact1}) as a consequence of  the negative sign of the kinetic term in corresponding effective Lagrangian \cite{KS}. 
This unphysical ghost   field  does not violate unitarity or any other important 
properties of the theory as  consequence  of Gupta-Bleuler-like condition on the physical Hilbert space  \cite{KS, Zhitnitsky:2011tr}.  This   description in terms of the KS ghost 
 implicitly  takes into account the presence  of  topological sectors in the system. The same property  is explicitly reflected 
 by summation over topological sectors $ { k \in \mathbb{Z}}$ in direct  computations  (\ref{Z_3},\ref{Z_4}) without introducing any auxiliary fields. 
 
 It is interesting to note that a similar structure also emerges  in other gauge theories, e.g. in the so-called ``deformed QCD". In that case  the topological sectors also produce the $\delta (x)$ function behaviour for the topological susceptibility, similar to eq. (\ref{local}).  Furthermore, this contact term in that model can be also described by a ghost  which turns out to be an auxiliary topological field  described  by the Chern-Simons topological action~\cite{Zhitnitsky:2013hs}.
 
 In next section we shall see that a similar contact term with $\delta (x)$ function behaviour also emerges in 4d Maxwell system defined on a compact manifold. Furthermore, this contact term in 4d system can be understood in terms of auxiliary topological fields in BF formulation as discussed in section \ref{BF}. The corresponding auxiliary non-propagating fields play the same role as KS ghost fields in 2d QED \cite{KS, Zhitnitsky:2011tr} and topological Chern-Simons fields in ``deformed QCD" as presented in ~\cite{Zhitnitsky:2013hs}.

 \subsection{Entropy in 2d Maxwell system }\label{2d-entanglement} 
 The partition function (\ref{Z_2}),  (\ref{Z_4}), (\ref{poisson})   computed above allows us to compute the entropy of the system. However, 
 before we proceed, we want to make a short historical detour on the entropy studies  in this ``empty" model.  
   
  It has been claimed \cite{Kabat:1995eq} that, using the so-called conical method,  the black hole entropy is equal to the entropy of entanglement for spins zero and one-half fields (at least at one loop level). However,  for  gauge  Maxwell field, the entropy has an extra term describing the contact interaction with the horizon.  While the entropy is a positively defined entity, the Kabat contact term   is negative \cite{Kabat:1995eq}. Furthermore, this term being a total divergence can be represented as a surface integral determined by the behaviour of the theory at arbitrarily large distances, i.e. it obviously has an infrared (IR) origin. More recently, it has been conjectured  \cite{Zhitnitsky:2011tr} that the  Kabat contact term is originated from the same topological gauge sectors 
  and tunnelling transitions which can  not be associated with any physical degrees of freedom 
  as there are none in the ``empty" 2d theory. 
   Next step in this development was explicit demonstration  \cite{Donnelly:2012st} that   the computation   of the entropy in    the 2d Maxwell system is highly sensitive to the IR physics. 
   Therefore, 
    the IR regularization  should be    treated very carefully.
  Appropriate treatment has been suggested  in  \cite{Donnelly:2012st}  by defining  the system in a large box size $V$. With this regularization the computation 
    for the entropy can be easily performed as the corresponding partition function for the 2d Maxwell system defined in a box is known (\ref{Z_2}), (\ref{Z_4}), (\ref{poisson}). The  conical entropy with this regularization coincides with  conventional thermodynamical entropy since the volume of the Euclidean manifold is liner in the deficit angle,    
    \be
   \label{entropy-definition}
   S(V)=\frac{\partial}{\partial T}(T\ln {\cal{Z}}) =
    \left(1-V\frac{\partial}{\partial V}\right)\ln {\cal{Z}}( V), 
   \ee
  such that the final expression for $S(V)$  can be represented in  the following form 
  \cite{Donnelly:2012st}  (see also \cite{Solodukhin:2012jh,Kabat:2012ns} with related discussions):
      \be
  \label{entropy}
  S(V)=\left(\ln {\cal{Z}}( V)+\frac{1}{2}\right)-\frac{1}{2}\left(\frac{4\pi^2}{e^2}\right)\chi (V).
   \ee
  In this formula $\chi (V)$ is the topological susceptibility  (\ref{chi1})  and ${\cal{Z}}(V)$  is the partition function  determined by (\ref{Z_2}), (\ref{Z_4}), (\ref{poisson}) at $\theta=0$, see next subsection \ref{2d-theta} for generalization on $\theta\neq 0$. 
  
  Few comments are in order. First, the entropy $S(V)\rightarrow 0$  approaches zero  in large volume limit $V\rightarrow\infty$ as it should as there are no any propagating degrees of freedom  present in the system. However, entropy vanishes in a quite  nontrivial way: a conventional contribution from (\ref{entropy}) $\sim \ln {\cal{Z}} $ approaches zero as ${\cal{Z}}\rightarrow 1$ according to (\ref{Z_2}).  At the same time,    the negative   contribution due to the topological term  (which is determined by its asymptotic value   $\chi\rightarrow{e^2}/{4\pi^2}$, see eq. (\ref{exact1})),  exactly cancels 
with the the positive contribution $1/2$ originated from the quantum fluctuations.   Second, 
    one can explicitly see from  (\ref{entropy})  that  the  gauge invariant negative contribution is indeed present in  this expression  for the entropy.   This term with the ``wrong" sign in   eq.(\ref{entropy}) is exactly proportional to the  topological susceptibility (\ref{chi1})  in agreement with conjecture \cite{Zhitnitsky:2011tr}.  
Third, this term    can be represented as   a surface integral because  $Q=\frac{e}{2\pi}E$ entering (\ref{chi1}) is the topological charge density operator which is a total divergence. Fourth,  while the  term $\sim \chi( V)$ in eq. (\ref{entropy}) can be represented as a surface integral, the entropy itself does not  possesses    such  a surface representation. Nevertheless, both these  entities ($S$ and $\chi$ ) are separately   gauge invariant observables.

Furthermore, the entropy (\ref{entropy})  is obviously  a positively defined function at any finite $V$ and can be interpreted as the entanglement entropy \cite{Donnelly:2012st}.  Indeed,  the only local observable is $E$, which is constant over space.  It implies that the measurements of $E$ will be perfectly correlated on the opposite sides of the system. We interpret the same feature of entanglement in a different way. Our interpretation is 
based on Euclidean formulation of the system when the partition function (\ref{Z_3}), (\ref{Z_4}) can be interpreted  as a probability  of tunnelling events between different topological sectors $|k\ra$ in volume $V$. 
Once a tunnelling event happens, the corresponding boundary conditions of the gauge filed are entangled   on opposite sides of the system  as topological charge (\ref{Q}) is unambiguously 
determined by these boundary conditions.
 
 Finally, we want to emphasize that the presence of the topologically protected term  in   eq.(\ref{entropy}) proportional to the  topological susceptibility (\ref{chi1}) is correlated with the fact that this ``empty" system, in fact,  is  the topological quantum field theory. We shall see that a similar correlation   also  holds for  4d Maxwell system   which also shows  a number of other manifestations  being  typical for topologically ordered systems. 
 
$\bullet$ We   conclude this subsection with the following  short comment on terminology. The term ``entanglement entropy" is normally used to describe the entangled properties of physical propagating degrees of freedom. Our entropy (\ref{entropy}) in ``empty" theory has fundamentally different nature as it does not correspond to any propagating states. Rather, 
entropy (\ref{entropy}) results from  topologically different but physically identical quantum winding states $|k\ra$ and their   dynamics  (tunnelling transitions between them).  Therefore,  it is more appropriate to coin this  entropy as ``homotopical entropy"  as corresponding   expression (\ref{entropy})  is in fact 
a direct consequence   of nontrivial homotopy of the gauge group $\pi_1[U(1)]= \mathbb{Z}$ in $2d$ QED. It can only emerge in gauge systems
with nontrivial topological features. In particular, it can not occur  in scalar field theories. To simplify terminology   we shall  refer to  (\ref{entropy}) as topological entropy\footnote{We define the TE as conventional thermodynamical entropy (\ref{entropy-definition}) with the only difference is  that the partition function $\cal{Z}$ in (\ref{entropy-definition}) describes the dynamics of degenerate winding states rather than the dynamics of real degrees of freedom. In 2d QED this is the only dynamics which is present in the system such that TE identically coincides with the conventional thermodynamical entropy. In four dimensional QED discussed in next section the conventional contribution $\sim {\cal{Z}}_{0}$ due to physical photons decouple from the topological contribution $\cal{Z}_{\rm top}$, i.e. ${\cal{Z}}={\cal{Z}}_{0}\times \cal{Z}_{\rm top}$. Therefore, the TE derived from $\cal{Z}_{\rm top}$ normally represents a small correction to conventional thermodynamical entropy (which is typically  proportional to the volume of the system) derived from ${\cal{Z}}_{0}$. However, the main point of this work is that  while  the TE is a parametrically small portion  in terms of a magnitude, it produces  a  topologically protected contribution to the entropy, similar to analogous feature of the  topological entanglement entropy  computed for  topologically ordered  condensed matter systems, see section \ref{TEE} with more comments.} (TE). Our ultimate goal of this work is to compute the TE in  four dimensional QED defined on 4-torus, similar to eq. (\ref{entropy}) describing $2d$ QED. As we shall see in section \ref{4d}  the corresponding properties of the TE   are fundamentally different from conventional thermodynamical entropy. In fact, the TE   more resembles 
the topological entanglement entropy introduced   in condensed matter literature to study the topologically ordered systems rather than
conventional thermodynamical entropy describing  propagating degrees of freedom.

  \subsection{Topological Entropy at $\theta\neq 0$ }\label{2d-theta} 
  Now we want to generalize the results of  section \ref{2d-entanglement} to include  $\theta\neq 0$.
  We want to see how the system varies  when  $\theta\neq 0$  and how the topological entropy reflects the corresponding variations.  This generalization for $\theta\neq 0$ will play an important  role in our discussions of 4d Maxwell system  in section \ref{4d}. 
  
  First of all, it has been known for many years   \cite{Print-76-0357 (HARVARD)} that a non-vanishing    $\theta\neq 0$ is equivalent to the uniform electric field present in the system. The simplest way to see this in the Hamiltonian approach  is to observe that $\theta\neq 0$ produces the shift of the electric field
  $n\rightarrow  \left(n+{\theta}/{2\pi}\right)$ in formula (\ref{Z_2}), i.e. a non-vanishing $\theta$ corresponds to the background electric field
  \be
  \label{E1}
  \la E\ra_{\rm Mink}= \frac{e\theta}{2\pi}.
  \ee
  The same result can be easily reproduced in Euclidean path integral approach   
  by differentiating partition function ${\cal{Z}}(\theta, V)$ with respect to $\theta$ according to the definition (\ref{Z_3}),  (\ref{theta}). Assuming that $|\theta| < \pi$ ($\theta=\pi$ requires a special treatment, and will be presented at the end of the section) we get: 
  \be 
  \label{E2}
 \la iQ\ra_{\rm Eucl}= \la \frac{ie E}{2\pi} \ra_{\rm Eucl}
  =\frac{\partial \ln {\cal{Z}}(\theta, V)}{V\partial \theta}|_{V\rightarrow\infty}  = -\frac{\theta e^2}{4\pi^2},~~~~
  \ee
  which coincides with (\ref{E1}) when written in Minkowski notations. 
   As a consequence of   nonzero background field (\ref{E1}, \ref{E2})  in the system the expression  for the topological entropy for $\theta\neq 0$  is slightly modified in comparison with (\ref{entropy}). Now it can be written in the following  form
      \be
  \label{entropy1}
  S(\theta, V)&=&\left(\ln {\cal{Z}}( \theta, V)+\frac{1}{2}\right)-\frac{1}{2}\left(\frac{4\pi^2}{e^2}\right)\chi (\theta, V)\nonumber \\
  &-&\frac{1}{2}\left(\frac{4\pi^2}{e^2}\right)\cdot V\cdot\la Q\ra \cdot\la Q\ra 
   \ee
  where new term appears as a result of non-vanishing background field. In this formula     $\la Q\ra$ and $\chi$
  are defined as before, see eqs. (\ref{exact1}) and  (\ref{E2}) in terms of first and second derivatives of  $\ln {\cal{Z}}(\theta)$ with respect to $\theta$ correspondingly, i.e.
  \be
\label{definition}
\la iQ\ra =\frac{1}{V }  \frac{\partial \ln {\cal{Z}}(\theta)}{\partial \theta}, ~~~
\chi (\theta,V)= -\frac{1}{V } \frac{\partial^2 {\ln\cal{Z}}(\theta)}{\partial\theta^2}. ~~~
\ee 
The topological entropy (\ref{entropy1}) approaches zero  in large volume limit as it should. But this vanishing result  is realized quite differently when the background field (\ref{E2}) is present in the system. Indeed, the partition function (\ref{Z_2})  in large volume limit, up to exponentially small corrections,  can be approximated as 
 \be
 \label{Z_5}
  \ln {\cal{Z}}(\theta, V\rightarrow \infty)\simeq  -  \frac{e^2V\theta^2}{8\pi^2} , 
 \ee
  which leads to the following asymptotic expressions for background field and topological susceptibility
  after differentiation according to definitions (\ref{definition}),   
   \be 
  \label{E3}
 \la Q\ra_{\rm Eucl} (\theta, V\rightarrow \infty)=i\frac{\theta e^2}{4\pi^2},\nonumber\\
 \chi(\theta, V\rightarrow \infty) =\frac{e^2}{4\pi^2}.
  \ee
By substituting (\ref{Z_5}) and (\ref{E3}) to eq. (\ref{entropy1}) one can verify that the entropy indeed vanishes (with exception of a single degenerate point $\theta=\pi +2\pi n$, see below) in infinite volume limit at arbitrary $\theta$ as it should.   The entropy $S(\theta, V)$ approaches zero from above in this case as a result of cancellation between two terms proportional to the volume: conventional $ \ln {\cal{Z}}$ contribution (\ref{Z_5}) cancels the  background field contribution which is also proportional to the volume (\ref{entropy1}). It is instructive to represent the same formula (\ref{entropy1}) in somewhat different way
  \be
  \label{entropy2}
  &&S(\theta, V)=\left(\ln {\cal{Z}}( \theta, V)+\frac{1}{2}\right)\\ \nonumber &-&\frac{1}{2}\left(\frac{4\pi^2}{e^2}\right)  
 \int d^2x \la Q^{\rm tot}(x), Q^{\rm tot}(0)\ra   ,
 \ee
where $Q^{\rm tot}=\la Q\ra+Q$ represents the total topological density operator including its background   portion (\ref{E2}) proportional to the constant electric field. One should note that the correlation function which appears in (\ref{entropy2}) can not be represented as second derivative of $\ln{\cal{Z}}(\theta)$ with respect to the $\theta$ 
as  defined by eq.(\ref{definition}). Instead, this term can be represented  as second derivative of ${\cal{Z}}$ itself, 
 \be
 \int d^2x \la Q^{\rm tot}(x), Q^{\rm tot}(0)\ra=-\frac{1}{V{\cal{Z}}(\theta)} \frac{\partial^2 {\cal{Z}}(\theta)}{\partial\theta^2}. 
      \ee

To conclude this section we want  to mention that 
the physics is perfectly $2\pi$ periodic with respect to $\theta$ as partition function (\ref{Z_2}, \ref{Z_4}) is obviously a $2\pi$ periodic function. 
In particular $\theta=2\pi$ is identically  the same as $\theta=0$ state. At the same time at $\theta=\pi$ 
the  double degenerate states appear in the system. This time this degeneracy corresponds to  physical degeneracy when two distinct vacuum states have equal energies.    One can see the emergence of this degeneracy by approaching $\theta=\pi$ point from opposite sides,  i.e. $\theta=\pi\pm\epsilon$ with $\epsilon\rightarrow 0$. These two degenerate states are classified by 
different directions of the electric field characterizing the system as we computed above, 
 \be
  \label{E4}
\la Q\ra_{\rm Eucl} =\pm\frac{i\pi e^2}{4\pi^2}, ~~  \la E\ra_{\rm Mink}=\pm \frac{e\pi}{2\pi}=\pm\frac{e}{2}.
  \ee
One should mention that similar formulae  for degenerate states with $\theta=\pi$  in this model were also recently discussed in \cite{ChenLee} in context of topological insulators. 

The expression for topological entropy (\ref{entropy1}) receives a crucial modification as partition function (\ref{Z_2})  has two identical terms for $\theta=\pi$ which correspond to the physical  degeneracy mentioned above. Indeed, in large volume limit the partition function at $\theta=\pi\pm\epsilon$ can be approximated as follows
   \be
 \label{theta_pi_2d}
  {\cal{Z}}(\theta=\pi\pm\epsilon)\simeq \left[e^{-\frac{e^2V}{2} \left( \frac{\pi \pm\epsilon}{2\pi}\right)^2} +e^{-\frac{e^2V}{2} \left(1- \frac{\pi\pm\epsilon }{2\pi}\right)^2} \right], ~~
     \ee
  such that 
  two terms contribute with equal weight at $\theta=\pi $. 
  As a consequence, the entropy  which can be easily computed by substituting the corresponding expressions to general formula (\ref{entropy1})  does not vanish at $\theta=\pi$, but rather   assumes 
  the following value
     \be 
  \label{entropy_pi}
  S(\theta=\pi)= \ln 2 .
  \ee
     The interpretation of this result is amazingly simple and straightforward. While our system is indeed ``empty" and  does not support any propagating degrees of freedom in the bulk, the   ground state is in fact   degenerate, and it is  
 characterized   by the vacuum expectation value of electric field (\ref{E4}).  
 Furthermore,  this degeneracy       which leads to extra term in the topological entropy (\ref{entropy_pi}) is, in fact, a volume independent phenomenon. Indeed, one can easily see that every term in the partition function (\ref{Z_2}) has its identical partner at $\theta=\pi$ at arbitrary volume $V$, which eventually leads to extra factor $\ln 2 $ in expression (\ref{entropy_pi}).  Therefore, the  emergence of non-vanishing entropy (\ref{entropy_pi}) in this ``empty" system is obviously a pure topological effect which reflects the two-fold vacuum degeneracy (\ref{E4}) present in the system. 
 Conventional thermodynamical entropy must vanish at $T=0$ which is obviously not the case for TE in thermodynamical limit at zero temperature given by eq.(\ref{entropy_pi}).

  \section{  Topological  Entropy  in Maxwell theory in four dimensions}\label{4d}
 The main goal  of this section is to derive formulae for  the  topological  entropy 
 in   Maxwell theory defined on 4d torus, similar to (\ref{entropy}), (\ref{entropy1}) derived for 2d system.
 The corresponding expressions will be entirely due to the tunnelling events similar to analysis 
 of ``empty" 2d system. The interpretation of the corresponding formulae will be also very similar to our  
 discussions of the 2d system. Namely, we will interpret the corresponding corrections to the entropy as topological  entropy, similar to our discussions in section \ref{2d-entanglement}, because the relevant topological configurations describe the tunnelling events   are uniquely determined by the properties of the topological configurations, similar to 2d analysis.  
 
 The crucial difference with 2d studies is, of course, that 
 Maxwell theory in four dimensions describes real physical photons with two transverse polarizations
 in contrast with our  studies of ``empty" 2d theory in section \ref{topology}. However, as we discuss below the propagating degrees of freedom with non-zero momentum completely decouple from topological contributions such that the partition function can be represented in the form ${\cal{Z}} = {\cal{Z}}_{0} \times {\cal{Z}}_{\rm top}$. 	
 
  The conventional,  topologically trivial   portion of entropy related to ${\cal{Z}}_{0}$ is well-known for  the Maxwell system.
  It is an extensive entity and it produces the vanishing entropy for zero temperature. The entanglement entropy for Maxwell theory  is also known,  see recent paper \cite{Solodukhin:2012jh} with many references on previous works therein. The corresponding contribution to the entanglement entropy is entirely determined by two transverse photon's polarizations and proportional to the area, similar to   scalar field theories, 
  \exclude{ In order to calculate the entanglement entropy one usually uses the so-called replica trick, see original papers \cite{Kabat:1994vj,Callan:1994py,Jacobson:1999mi} and review \cite{Solodukhin:2011gn}.
 Conventional procedure consists of introduction of  a small angle deficit at the surface  and differentiation of  the obtained result with respect to the angle deficit. This procedure normally leads to expected entanglement entropy, see recent \cite{Solodukhin:2012jh} and references therein,  }
  \be
  \label{entropy-area}
  S\sim A(\Sigma)(d-2)
  \ee
  where $A(\Sigma)$ is the area of surface $\Sigma$ and $(d-2)$ is the number of the on-shell physical propagating degrees of freedom in $d$ dimensional space. This leading 
   term is related to physical propagating degrees of freedom.  The  conventional thermodynamical entropy $S\sim VT^3$ and  entanglement entropy  (\ref {entropy-area}) are {\it not } subject of  the present work as they   completely decouple  from topological contributions related to ${\cal{Z}}_{\rm top}$, which is the main subject of our present studies. To avoid confusion with terminology we also emphasize that 
   these entropies  (conventional thermodynamical entropy as well as entanglement entropy)  do not depend on topological $|k\ra$ sectors of the system  due to linearity of the Maxwell theory, see details below. Therefore, they can not depend on $|k\ra$, nor they can carry any information about topological features of the system. 
   
 In the rest of this section we will concentrate on    behaviour of ${\cal{Z}}_{\rm top}$ and the expression for TE  which follows from ${\cal{Z}}_{\rm top}$.   In other words, our goal here is to study the corrections to the  thermodynamical entropy   due to topological configurations describing the tunnelling events, rather than contributions related to the physical propagating photons, similar to our analysis  of the ``empty" theory in section \ref{topology}. For these computations  we use the same definition (\ref{entropy-definition}) for the thermodynamical entropy we have been using before in our studies of the two dimensional system.    
  
\subsection{Topological partition function ${\cal{Z}}_{\rm top}$.}\label{Ztop}
 Our goal here is to define the Maxwell system   on a Euclidean 4-torus   with  sizes $L_1 \times L_2 \times L_3 \times \beta$ in the respective directions. It provides the IR regularization of the system. As we discussed  in section   \ref{2d-entanglement} this IR regularization plays a key role in proper treatment of the contact topological term which is related to tunnelling events rather than   the propagation of the  physical photons with transverse polarizations.

 We follow \cite{Cao:2013na} in our construction of the partition function ${\cal{Z}}_{\rm top}$ where it was employed    for  computation of  the corrections to the Casimir effect due to these novel type of topological fluctuations. The crucial point is that we impose the periodic boundary conditions on gauge $A^{\mu}$ field up to a large gauge transformation.
 In what follows we simplify our analysis by considering   a clear case with winding topological sectors $|k\ra$    in the z-direction only.  The classical configuration in Euclidean space  which describes the corresponding tunnelling transitions can be represented as follows:
\be
\label{topB4d}
\vec{B}_{\rm top} &=& \vec{\nabla} \times \vec{A}_{\rm top} = \left(0 ,~ 0,~ \frac{2 \pi k}{e L_{1} L_{2}} \right),\\
\Phi&=&e\int dx_1dx_2  {B}_{\rm top}^z={2\pi}k \nonumber
\ee
in close analogy with the 2d case (\ref{Q}).

The Euclidean action of the system is quadratic and has the following  form  
\be
\label{action4d}
\frac{1}{2} \int \dd^4 x \left\{  \vec{E}^2 +  \left(\vec{B} + \vec{B}_{\rm top}\right)^2 \right\} ,
\ee
where $\vec{E}$ and $\vec{B}$ are the dynamical quantum fluctuations of the gauge field.  
The key point is that the classical topological portion of the action decouples from quantum fluctuations, such that the quantum fluctuations do not depend on topological sector $k$ and can be computed in topologically trivial sector $k=0$.
Indeed,  the cross term 
\be
\int \dd^4 x~ \vec{B} \cdot \vec{B}_{\rm top} = \frac{2 \pi k}{e L_{1} L_{2}} \int \dd^4 x~ B_{z} = 0 
\label{decouple}
\ee
vanishes  because the magnetic portion of quantum fluctuations in the $z$-direction, represented by $B_{z} = \partial_{x} A_{y}  - \partial_{y} A_{x} $, is a periodic function as   $\vec{A} $ is periodic over the domain of integration. 
This technical remark in fact greatly simplifies our  analysis as the contribution of the physical propagating photons 
is not sensitive to the topological sectors $k$. This is,  of course,  a specific feature  of quadratic action 
 (\ref{action4d}), in contrast with non-abelian  and non-linear gauge field theories where quantum fluctuations of course depend on topological $k$ sectors. The authors of 
 ref. \cite{ChenLee} arrived to the same   conclusion (on decoupling  of the  topological terms from  conventional fluctuating photons with non-zero momentum),   though in a different context of topological insulators in the presence of the $\theta=\pi$ term. 
 
The classical action  for configuration (\ref{topB4d}) takes the form 
\be
\label{action4d2}
\frac{1}{2}\int \dd^4 x \vec{B}_{\rm top}^2= \frac{2\pi^2 k^2 \beta L_3}{e^2 L_1 L_2}
\ee
To simplify our analysis further in  computing  ${\cal{Z}}_{\rm top}$ we consider a geometry where $L_1, L_2 \gg L_3 , \beta$ similar to construction relevant for the Casimir effect  \cite{Cao:2013na}. 
 In this case our system   is closely related to 2d Maxwell theory by dimensional reduction: taking a slice of the 4d system in the $xy$-plane will yield precisely the topological features of the 2d torus considered in section \ref{topology}.  
 Furthermore, with this geometry our simplification (\ref{topB4d}) when we consider exclusively the magnetic fluxes in $z$ direction is justified as the corresponding classical action (\ref{action4d2}) assumes a minimal  possible values\footnote{\label{e-flux}There are also electric fluxes $\Phi_E$ in the system in description of the Euclidean path integral. The corresponding electric fluxes are originated from the requirement that the electromagnetic  potential $\vec{A}$ satisfies the periodic boundary conditions up to the large gauge transformations, i.e. $A_3(\beta)=A_3(0)+\frac{2\pi l}{eL_3}$ which corresponds to the electric flux with uniform electric field $E_3=\frac{2\pi l}{e\beta L_3}$. The  Euclidean action for corresponding configurations is parametrically $({L_1L_2}/{\beta L_3}) $ larger than the magnetic classical action (\ref{action4d2}), and therefore it is consistently neglected in our analysis.  One should also note that the electric field in the  Euclidean classical action   must not be confused with electric field in the Hamiltonian formulation where it is a constant of motion, see e.g.\cite{Hermele:2004zz,ChenLee} where the electric fluxes emerge in Hamiltonian description in quite different context (the $U(1)$ spin liquid and topological insulators correspondingly).   The corresponding electric fields in these two descriptions (Euclidean path integral approach vs Hamiltonian approach) are in fact related by the duality transformation, similar to 2d analysis in section \ref{topology}  where electric field in Hamiltonian formulation enters formula (\ref{Z_2}), while electric flux in Euclidean path integral formulation enters eq. (\ref{Q},\ref{Z_4}), see  \cite{Cao:2013na}  for more details and references.}. With this assumption we can consider very small temperature, but still we can not take a formal limit $\beta\rightarrow\infty$  in our final expressions
 as a result of our technical constraints in the system. 
      
With these additional simplifications   the topological partition function becomes \cite{Cao:2013na}:
\be
\label{Z4d}
{\cal{Z}}_{\rm top} = \sqrt{\frac{2\pi \beta L_3}{e^2 L_1 L_2}} \sum_{k\in \mathbb{Z}} e^{-\frac{2\pi^2 k^2 \beta L_3}{e^2 L_1 L_2} },
\ee
 which is essentially the dimensionally reduced expression for  the topological partition function (\ref{Z_4}) for 2d 
 Maxwell theory analyzed in section \ref{topology}. 
   
One should note that the dimensional reduction which is employed here is not the most generic one.
In fact, one can impose a non-trivial boundary condition on every slice in the 4d torus, see comments in footnote \ref{e-flux} and ref.\cite{Kelnhofer:2012ig} for most generic construction. However, the main goal of this work is not   a generic classification. Rather, we wish  to discuss the  contact term, topological entropy, degeneracy and other nontrivial features by considering  the simplest possible setup  (\ref{topB4d}) when physics can be easily understood and analyzed. In other words, we wish to consider    a nontrivial BC  imposed on a single slice, while keeping  the  trivial periodic BC for other slices. 
  
We follow \cite{Cao:2013na} and  introduce the dimensionless parameter
\be
\label{tau}
\tau \equiv {2 \beta L_3}/{e^2 L_1 L_2}
\ee
such that the partition function ${\cal{Z}}_{\rm top}$ can be written in the dual  form:
\be
\label{Z_top}
{\cal{Z}}_{\rm top} (\tau)= \sqrt{\pi \tau} \sum_{k\in \mathbb{Z}} e^{-\pi^2 \tau k^2} = \sum_{n\in \mathbb{Z}} e^{-\frac{n^2}{\tau}}, 
\ee
where  the Poisson summation formula (\ref{poisson}) is used again.
Our normalization  of the partition function ${\cal{Z}}_{\rm top}$ 
is such that in the limit $L_1L_2\rightarrow \infty~ (\tau\rightarrow 0)$  the topological portion of the partition function ${\cal{Z}}_{\rm top}
\rightarrow 1$ as one can see from the dual representation (\ref{Z_top}). In this limit   the dimensional reduction is justified and  we  
 recover the conventional  physics  which is encoded in   ${\cal{Z}}_{0}$. This is  a result of the same decoupling of the topological transitions from  physics related to propagating photons, as we  discussed above when ${\cal{Z}} ={\cal{Z}}_{0}\times {\cal{Z}}_{\rm top}$  and ${\cal{Z}}_{\rm top}
\rightarrow 1$ in this limit. 
One should note that the normalization factor $\sqrt{\pi \tau}$ which appears in eq. (\ref{Z4d}) does not depend on topological sector $k$, and essentially it represents our convention of the  normalization   ${\cal{Z}}_{\rm top}
\rightarrow 1$ in the limit $L_1L_2\rightarrow \infty$.  
   
\subsection{Topological Entropy   and Magnetic Susceptibility in 4d Maxwell System} 
We are in position now to compute the TE associated with ${\cal{Z}}_{\rm top}$. As we mentioned previously, we shall use the same definition for the entropy we used previously in 2d studies  (\ref{entropy-definition}),  i.e.
 \be
   \label{entropy4}
   &&S_{\rm top}(\tau)=     \left(1-\beta\frac{\partial}{\partial \beta}\right)\ln {\cal{Z}}_{\rm top}(\tau)\\
   &=& \ln {\cal{Z}}_{\rm top}(\tau)-\frac{1}{\tau}\cdot\frac{1}{{\cal{Z}}_{\rm top}(\tau)} \cdot\sum_{n\in \mathbb{Z}} n^2 e^{-\frac{n^2}{\tau}} , \nonumber
   \ee
where we use the dual representation (\ref{Z_top}) for the partition function ${\cal{Z}}_{\rm top}(\tau)$.  

Our next step is to represent the second term in (\ref{entropy4}) in a form, similar to expression for the entropy (\ref{entropy}) in terms of the topological susceptibility in two dimensional theory. To achieve this goal we formally introduce $\theta_{\rm eff}$ into the dual representation (\ref{Z_top}) for the partition function  
as follows
\be
\label{Z_dual}
{\cal{Z}}_{\rm top} (\tau, \theta_{\rm eff}) = \sum_{n\in \mathbb{Z}} e^{-\frac{n^2}{\tau}+in\theta_{\rm eff}}. 
\ee
One should emphasize that the $\theta_{\rm eff}$ parameter introduced in (\ref{Z_dual}) is not a fundamental $\theta$ parameter normally introduced into the Lagrangian  in front of  $\vec{E}\cdot\vec{B}$ operator. Furthermore, integer 
number $n$ which appears in front of  $\theta_{\rm eff}$   in (\ref{Z_dual}) is not the magnetic flux $k$ defined by eq. 
(\ref{topB4d}) which enters original partition function (\ref{Z4d}).  Rather, integer $n$ appears in the dual form (\ref{Z_top}) for this partition function. Nevertheless, as we discuss in next section \ref{magnetic} the parameter $\theta_{\rm eff}$ has a perfect physical meaning related to the   external magnetic flux through the $xy$-plane applied to the system, 
\be
\label{theta_eff}
\theta_{\rm eff} = B^{\rm ext}_z L_1L_2e.
\ee
Now, the second term in (\ref{entropy4}) can be formally represented as the second derivative of ${\cal{Z}}_{\rm top} (\tau, \theta_{\rm eff})$ with respect to $\theta_{\rm eff}$. Indeed, using   identity
 \be
   \label{entropy5}
   \frac{1}{{\cal{Z}}_{\rm top}(\tau)} \cdot\sum_{n\in \mathbb{Z}} n^2 e^{-\frac{n^2}{\tau}} =-
   \frac{\partial^2 \ln {\cal{Z}}_{\rm top}(\tau, \theta_{\rm eff})}{\partial \theta_{\rm eff}^2}|_{\theta_{\rm eff}=0}~~~
   \ee
 one can rewrite  the expression for entropy (\ref{entropy4})   in the following    form  
 \be
   \label{entropy6}
    S_{\rm top}(\tau)   
   &=& \ln {\cal{Z}}_{\rm top}(\tau) -\frac{1}{2}\chi_{\rm mag}(\tau), \\
   \chi_{\rm mag}(\tau)&\equiv& -\frac{2}{\tau}
\frac{\partial^2 \ln {\cal{Z}}_{\rm top}(\tau,\theta_{\rm eff})}{\partial \theta_{\rm eff}^2}|_{\theta_{\rm eff}=0}.
\nonumber
      \ee
Significance of this representation is that   $\chi_{\rm mag}(\tau)$ 
  entering the expression (\ref{entropy6}) has many features similar to the topological susceptibility 
entering (\ref{entropy})  in two dimensional theory. In particular, $\chi_{\rm mag}(\tau)$ can be represented as  a surface integral and  it assumes a  finite value   in the thermodynamical limit. Furthermore,  $\chi_{\rm mag}(\tau)$   is not sensitive to any specific details in the bulk of the system, nor its boundary's geometrical shape. Rather it is only sensitive  to   the boundary conditions which globally classify the topological sectors of the system.

Furthermore, as we shall see in a moment  $\chi_{\rm mag}(\tau)$ is in fact the conventional magnetic susceptibility which measures response of the free energy to the introduction of arbitrary small external magnetic field. 
This is because the formal parameter  $\theta_{\rm eff}$ entering (\ref{Z_dual}) is related to the physical external field $B^{\rm ext}_z$ as  eq. (\ref{theta_eff}) states.     As a result of this relation, the differentiation of $\ln {\cal{Z}}_{\rm top}(\tau, \theta_{\rm eff})$ with respect to $\theta_{\rm eff}$ is equivalent to differentiation  with respect to external field $B^{\rm ext}_z$ which is, by definition, the conventional magnetic susceptibility, 
  \be 
  \label{chi_mag}
&&\chi_{\rm mag} (\tau) = -\frac{2}{\tau}
\frac{\partial^2 \ln {\cal{Z}}_{\rm top}(\tau, \theta_{\rm eff})}{\partial \theta_{\rm eff}^2}|_{\theta_{\rm eff}=0}\\ &=&
-\frac{1}{\beta V}\frac{\partial^2 \ln \mathcal{Z}_{\rm top}}{\partial B_{\rm ext}^2}|_{B_{\rm ext}=0}
=  \int d^4x \la B_z(x), B_z(0)\ra  .~~\nonumber 
\ee
 Representation (\ref{chi_mag}) for $\chi_{\rm mag} (\tau)$  entering the expression for the entropy (\ref{entropy6}) obviously implies that this term is a total divergence, as $B^i=\epsilon^{ijk}\partial_jA_k$.   In other words, 
 this topological contribution to the entropy is determined by the  behaviour of the gauge fields at arbitrary large distances,  
similar to our studies  of  the entropy in section \ref{topology} in 2d ``empty" theory with 
  relation (\ref{entropy})   being  a  precise analog of eq. (\ref{entropy6}). 

The crucial difference between these two cases is, of course, that   4d Maxwell system describes real physical massless photons, in contrast with ``empty" 2d theory. However, the topological contribution, which is main subject of the present work, behaves in Maxwell  theory formulated on 4d torus very much in the same way as in 2d case. Furthermore, the interpretation of these topological terms in 4d theory is also very much the same as in 2d case. To be more specific,   we interpret (\ref{entropy6}) as TE (which is a sub-leading contribution to the thermodynamical entropy) resulting from tunnelling processes.  This term is always much smaller than  the leading term   $S\sim VT^3$   originating from conventional propagating physical photons with two transverse polarizations. 

To get some feeling on numerical (un)importance  of the topological terms  (\ref{entropy6})  in comparison with conventional leading term    we consider  $\tau$ parameter defined by eq. (\ref{tau}) to be very large 
$\tau\gg 1$ assuming the thermodynamical limit at very low (but non-vanishing) temperature\footnote{one should note that  large $\tau\gg 1$ is consistent  with our ``technical" simplification related to the dimensional  reduction employed in (\ref{Z4d}). In particular,    $\tau\gg 1 $ can be always arranged by    considering very small $e\rightarrow 0$ in eq. (\ref{tau}) before considering the dimensional reduction employed in (\ref{Z4d}). Still, we can not put $\tau\sim \beta=\infty$ because our simplified computations based on dimensional reduction require $L_1L_2\gg \beta L_3$.}. In this case 
   \be
   \label{}
   {\cal{Z}}_{\rm top}(\tau\gg 1)\rightarrow\sqrt{\pi\tau},
   \ee
while the topological   entropy assumes  the following asymptotic value
 \be
   \label{entropy_4d}
    S_{\rm top}(\tau\gg 1)  \rightarrow \left[\frac{1}{2}\ln({\pi\tau})-\frac{1}{2}\right].
       \ee
    The magnetic susceptibility asymptotically approaches unity in this limit,
      \be
      \label{chi_4d}
       \chi_{\rm mag} (\tau\gg 1) \rightarrow 1.
\ee
It is very instructive to compare the behaviour (\ref{entropy_4d}) with similar formula (\ref{entropy}) in 2d case. In both cases the topologically protected non-dispersive contact contribution  (a contribution which can not be expressed in terms of physical propagating degrees of freedom)
 approaches one and the same constant ($-1/2$). In 2d case this contact term is  
 related to the topological susceptibility (\ref{exact1})  while in 4d case the contact terms is formulated in terms  of the magnetic susceptibility (\ref{chi_mag}). 
In the 2d case this term  cancels with another positive contribution 
as the  total entropy is  determined by one and the same   partition function   $ {\cal{Z}} $, see section \ref{2d-entanglement}. In the 4d case  the same factor ($-1/2$) remains untouched and stays in eq.(\ref{entropy_4d})
as it represents only the topological portion of the entropy, not the total entropy. One can argue that the total entropy in the limit $\beta\rightarrow \infty$ also vanishes \cite{Kelnhofer:2012ig}, similar to 2d case. However, in the 4d case 
  it vanishes as a result of cancellation of  the topological term  (\ref{entropy_4d}) with the conventional  contribution computed on the 4 torus and related to the   physical propagating   photons described by  $ {\cal{Z}}_0 $.

  One can argue that $ \chi_{\rm mag}$ from (\ref{chi_4d}) is saturated by a non-dispersive  contact term which can not be associated with any physical propagating degrees of freedom, similar to 2d expression (\ref{local}). To be more precise, the integrand for $ \chi_{\rm mag}$ is expected  to have the following structure 
  \be
      \label{mag_local}
     \la B_z(x), B_z(0)\ra =\frac{\delta^2(x)}{L_3\beta}  , ~~\chi_{\rm mag} = \int_{{\mathbb{T}}^4} \frac{\delta^2(x) d^4x}{L_3\beta} =1,~~~~~
\ee
  where $ \delta^2(x) $ should be understood as the discretized version of the  delta function defined on the torus
  \be
  \label{delta_1}
  \delta^2(x)= \frac{1}{L_1L_2} \sum_{n_1n_2}e^{2\pi i\left(\frac{n_1x_1}{L_1}+\frac{n_2x_2}{L_2}\right)}.
  \ee
  Our argument supporting $\delta^2(x) $ function behaviour in eq. (\ref{mag_local}) is based on observation that the magnetic susceptibility has non-dispersive nature. Indeed, $\chi_{\rm mag} $  is derived from topological partition function  ${\cal{Z}}_{\rm top}(\tau) $ which completely decouples from  ${\cal{Z}}_{0}$  describing   propagating photons  with physical transverse polarizations. Therefore, any non-vanishing correlation function, including (\ref{mag_local}) must be expressed in terms of a  contact term with  structure (\ref{mag_local}) similar to the  contact term (\ref{local}) in 2d system.   
  Explicit computations in terms of the auxiliary fields using the so called ``BF" formulation in section \ref{BF} also supports the  structure (\ref{mag_local}). Furthermore, one can argue that the non-dispersive  contact term (\ref{mag_local}) is related to the IR physics at   large  
  distances rather than UV physics,  in close analogy  to our discussions of the 2d case (\ref{divergence}).
     
  Does it make any sense to keep this sub-leading term  ($-1/2$)   in eq.(\ref{entropy_4d}) in the presence of much greater conventional contribution $S\sim VT^3$ and $\ln({\pi\tau})$ also entering (\ref{entropy_4d})?
  Our ultimate answer is ``yes" as this constant factor proportional to the magnetic susceptibility $- 1/2 \chi_{\rm mag}$  has some universal topological properties as we shall argue below.
  
\subsection{Similarities and differences between TE and Topological Entanglement   Entropy  in CM systems }\label{TEE}
Before we proceed with  our arguments we want to make a short detour on   Topological Entanglement Entropy   in 3d as well as in 4d cases in condensed matter (CM) systems\footnote{not to be confused  with conventional  CM notations, where  it is a customary to count the spatial number of dimensions, rather than total number of dimensions,  such that our 4d system corresponds to $(D+1)$ Maxwell theory with $D=3$ in CM notations.}.  The main purpose  for this detour is to present some analogies and  similarities between these two  very different entities: TE discussed in last section versus topological entanglement entropy  introduced in refs \cite{Kitaev:2005dm,Levin:2006zz}. 

 The entanglement entropy in arbitrary number of dimensions is proportional to the surface area of the boundary. In 4d case it corresponds  to the area law (\ref{entropy-area}). In 3d   system the leading term  is proportional to the length $L$. We shall not discuss this leading term in the present work. 
 A portion  of   the entanglement entropy   which is important for our discussions  is in fact a  sub-leading term which may emerge in some systems.  Well known example is a   3d dimensional system in a   topologically ordered  phase where the first sub-leading term is universal constant, the so-called topological entanglement   entropy   (TEE)  and independent  on the size or shape of $L$ as argued in refs \cite{Kitaev:2005dm,Levin:2006zz}, i.e.
\be
   \label{entropy_3d}
    S_{\rm } =aL -\ln {\cal D}
           \ee
where $a$ is a non-universal  and cutoff -dependent coefficient, while ${\cal D}$ is  the  so-called   total quantum dimension of the topological phase, and it is universal constant  \cite{Kitaev:2005dm,Levin:2006zz}.  In other words, any small variations of the  system do not change ${\cal D}$. It has been argued that the presence of such term is potentially very useful probe of a topological phase, see e.g. original papers \cite{Wen:1989iv,Wen:1990zza, Moore:1991ks, BF, Cho:2010rk} and recent  reviews~\cite{Wen:2012hm, Sachdev:2012dq, Grover:2013ifa}  with large number of original references therein. Similar studies in 4d had received much less attention in the past. Still, it is known that a constant term similar to $\ln {\cal D}$ in (\ref{entropy_3d}) can appear in a 4d system    even for a non-topologically ordered phase. However, when the system is defined on flat space-time (e.g. has the topology of a torus,  which is precisely our case) the constant term in the entropy would signal    that the system is in fact in topologically ordered phase~\cite{Grover:2011fa}. 

After this short detour we return to our Maxwell system in four dimensions.  The topological part of the  system characterized by partition function (\ref{Z4d}), (\ref{Z_top})  obviously describes some sort of entanglement,  similar to 2d ``empty" theory as discussed in section \ref{topology} and ref.\cite{Donnelly:2012st}. This is because the fluxes which saturate the partition function (\ref{Z4d}), (\ref{Z_top}) are constant over space, which means that the measurement of the field will be perfectly correlated on the opposite sides of the system, similar to arguments  of ref.\cite{Donnelly:2012st} presented for 2d ``empty" theory. However, this is not a conventional entanglement describing physical propagating degrees of freedom in CM systems. Rather,  our system is formulated in terms of ``instantons"  (instead of propagating quasiparticles in CM systems) in Euclidean space-time with action (\ref{action4d2}). These pseudo-particles  saturate the topological portion of the partition function (\ref{Z4d}), (\ref{Z_top}). Such topological fluctuations occur even when no propagating degrees of freedom  exist  in the system as 2d example from section \ref{topology}  shows. 

As a result of this difference we can not use many  standard tools which normally would  detect the topological order. For example, we can not compute 
 the braiding phases of charges and vortices 
which are normally used in CM systems simply because our system does not support    such kind of excitations.  
Furthermore, the ``degeneracy"  in our system
  is related to degenerate of winding states $| k\ra$ which are connected to each other by large gauge transformation, and therefore
 must be identified as they  correspond to   the same physical state.
     It is very different from  conventional term ``degeneracy" in  topologically ordered CM systems  when {\it distinct} degenerate states are present in the system as a result of formulation of a theory on a topologically non-trivial manifold such as torus.  Finally, our system supports conventional massless photons with physical polarizations, in contrast with conventional topologically ordered phases characterized by a gap. However, these massless degrees of freedom   completely decouple  from our topological fluctuations according to  eq. (\ref{decouple}).  Formally, this decoupling is  expressed as   ${\cal{Z}} ={\cal{Z}}_{0}\times {\cal{Z}}_{\rm top}$ as discussed after eq. (\ref{decouple}).
     Therefore, these massless physical photons can be completely ignored in our discussions  of the topological properties of the partition function ${\cal{Z}}_{\rm top}$. In this respect it is very similar to topologically ordered superconductors \cite{BF} when massless phonons always exist in the system, but nevertheless, they completely decouple from relevant dynamics,  and can be ignored in discussions of the topological features of the model.

  In spite of the differences mentioned above, it is very instructive to compare the topological portions of the  TE  given by coefficient $-1/2$ in eq. (\ref{entropy_4d}) for  Maxwell system and $-\ln {\cal D}$ in eq. (\ref{entropy_3d}) for  CM system. 
  In both cases these terms are topologically protected, i.e. they are	not	sensitive to any specific details in the bulk of the system, nor the boundary's geometrical shapes. Rather these terms are determined by the   global properties of the systems. Indeed, in case  of 4d Maxwell system  the coefficient $-1/2$ in eq. (\ref{entropy_4d}) is expressed in terms of the magnetic susceptibility (\ref{chi_4d}) which itself is represented by a surface integral, while in CM systems the topological protection was advocated in refs.\cite{Kitaev:2005dm,Levin:2006zz}. Furthermore, in both cases these topological contributions, being the sub-leading terms,  have a negative sign in comparison with the leading terms.
    
  Therefore, the TE behaves very much in the same way as topological entanglement entropy  does, though TE can not be interpreted 
 in terms of the quantum dimension $ {\cal D}$   entering (\ref{entropy_3d})  as in CM systems. This is    because, as we already mentioned, the  quasiparticles  which can propagate   simply do not exist    in this Euclidean Maxwell system. Nevertheless, if we   equalize $-\ln {\cal D}$ from  eq. (\ref{entropy_3d}) and $-1/2$ from  (\ref{entropy_4d}),   we arrive to a formal
  relation 
  \be
  \label{e}
  -\ln {\cal D}=-\frac{ \chi_{\rm mag} (\tau\gg 1 )}{2}   =-\frac{1}{2} ~\longrightarrow~{\cal D}=\sqrt{e},~~~
  \ee
  which should be compared with conventional ${\cal D}=\sqrt{m} $ for a Laughlin state in a fractional quantum Hall system with filling factor $\nu=1/m$, or ${\cal D}=2$ for $p+ip$ superconductor, or any other similar systems,  see recent reviews  \cite{Wen:2012hm,Sachdev:2012dq} for the details and original references. 
  The emergence of the exponential function $e^{1/2}$ in (\ref{e}) instead of $m^{1/2}$  hints that TE in our system is originated from tunnelling transitions rather than from  dynamics of quasiparticles in the system described by quantum dimensions ${\cal D}$. This interpretation  is obviously consistent with our construction  of the partition function  $ {\cal{Z}}_{\rm top}(\tau)$ describing the tunnelling events between topologically different but physically identical $|k\ra$ states. Still,  these different entities $ -\ln {\cal D}$ in eq. (\ref{entropy_3d}) and $-\frac{1}{2}\chi_{\rm mag} (\tau)$ in (\ref{entropy6})   in very different systems behave very similarly under small variations of the systems, which justifies our comparison in form of  equalizing these two  different things represented by eq. (\ref{e}).   
  
  The main message of this subsection is  as  follows. 
  \exclude{The topological contribution to the partition function representing the transitions between different topological sectors  $|k\ra$ is known (\ref{Z_top}).
  Therefore, computations of the corresponding thermodynamical parameters (including corrections to the entropy) does not represent any additional information in comparison with the original expression for $ {\cal{Z}}_{\rm top}(\tau)$.  However, the main    observation is that}
  We observe that  a  sub-leading correction $- {1}/{2}\chi_{\rm mag} (\tau)$ in (\ref{entropy6})  to thermodynamical entropy $S\sim VT^3$  is topologically protected in the same way as TEE is protected in CM systems (\ref{entropy_3d}). Therefore, this sub-leading term might be  signalling  that our system behaves  as a topologically ordered CM system\footnote{This is in spite of the fact that our system of course supports massless photons in contrast with fully gapped CM systems. However, as explained above the conventional massless degrees of freedom completely decouple from  topological contributions as eq. (\ref{decouple}) states.}. 
 \exclude{ In our specific case this  is, of course, the expected result as underlying construction of 
  $ {\cal{Z}}_{\rm top}(\tau)$ is formulated in terms of the topological objects (\ref{topB4d}) which  describe the interpolations between different topological $|k\ra$  sectors.   However, in more complicated  cases a partition function may not be known exactly, in which case a topologically protected contribution similar to (\ref{e}) may be  signalling the presence of some   topological features in a system, in close analogy with CM studies (\ref{entropy_3d}). }
  Furthermore,   our system     demonstrates a property of  physical degeneracy of the ground state 
    when $\theta_{\rm eff}=\pi$, similar to 2d case   (\ref{E4}), as we shall discuss below.
The emergence of such  degeneracy in a system    is a typical manifestation of a topological order in CM systems. In   next section \ref{BF} we will  reformulate   the same Maxwell system in terms of  the so-called ``BF" action.  A  similar ``BF"  structure in CM systems is known to  describe a large distance behaviour in a  topologically ordered phases. Therefore, such a BF representation of the Maxwell system in section \ref{BF} is an additional argument supporting our claim that the Maxwell system defined on a compact torus belongs to a topologically ordered phase. 
    
\subsection{Topological entropy in the background  of a magnetic field }\label{magnetic}
In this section we want to generalize our results on TE for Euclidean Maxwell system in the presence of the external magnetic field. Normally, in the conventional quantization of electromagnetic fields in Minkowski space, there is no \emph{direct} coupling    between fluctuating vacuum photons and an external magnetic field as a consequence of linearity of the Maxwell system. The coupling with  fermions   generates  a negligible effect $\sim \alpha^2B_{ext}^2/m_e^4$ as the non-linear Euler-Heisenberg Effective Lagrangian  suggests, see \cite{Cao:2013na} for the details.  The interaction of the external magnetic field with topological fluctuations (\ref{topB4d}),  in contrast with coupling with conventional   photons, will  lead to the effects of order of unity as a result of interference of the external magnetic field with fluxes- instantons.

The corresponding partition function can be easily constructed for external magnetic field $B_{z}^{\rm ext}$ pointing along $z$ direction, as 
the crucial technical element on decoupling of the background fields  from quantum fluctuations assumes the same form (\ref{decouple}).  In other words, the physical propagating photons with non-vanishing momenta are not sensitive to the topological $k$ sectors, nor to the external uniform magnetic field, similar to our discussions after (\ref{decouple}).

The classical action for configuration in the presence of the uniform external magnetic field $B_{z}^{\rm ext}$ therefore takes the form 
\be
\label{B_ext}
\frac{1}{2}\int \dd^4 x  \left(\vec{B}_{\rm ext} + \vec{B}_{\rm top}\right)^2=  \pi^2\tau\left(k+\frac{\theta_{\rm eff}}{2\pi} \right)^2
\ee
where $\tau$ is defined by (\ref{tau}) and  the effective theta parameter $\theta_{\rm eff}$ is expressed in terms of the original external magnetic field (\ref{theta_eff}).
Therefore, the partition function in the presence of the uniform magnetic field can be reconstructed from (\ref{Z4d}) and it is given by \cite{Cao:2013na}, 
\be 
\label{Z_eff}
  {\cal{Z}}_{\rm top}(\tau, \theta_{\rm eff})
 =\sqrt{\pi\tau} \sum_{k \in \mathbb{Z}} \exp\left[-\pi^2\tau \left(k+\frac{\theta_{\rm eff}}{2\pi}\right)^2\right].~~
\ee
  The dual representation for this partition function is obtained by applying the Poisson summation formula (\ref{poisson})
  \be 
\label{Z_dual1}
  {\cal{Z}}_{\rm top}(\tau, \theta_{\rm eff})
 &=&\sqrt{\pi\tau} \sum_{k \in \mathbb{Z}} \exp\left[-\pi^2\tau \left(k+\frac{\theta_{\rm eff}}{2\pi}\right)^2\right] \nonumber \\
  &=& \sum_{n\in \mathbb{Z}} \exp\left[-\frac{n^2}{\tau}+in\cdot\theta_{\rm eff}\right], 
  \ee
   which justifies our notation for  the effective theta parameter $\theta_{\rm eff}$ as it enters the partition function in combination with integer number $n$. One should emphasize that integer  number $n$ in the dual representation (\ref{Z_dual1}) is not the integer magnetic flux $k$ defined by eq. 
(\ref{topB4d}) which enters original partition function (\ref{Z4d}). Furthermore,  the $\theta_{\rm eff}$ parameter which enters (\ref{Z_eff}, \ref{Z_dual1}) is not a fundamental $\theta$ parameter which is normally introduced into the Lagrangian  in front of  $\vec{E}\cdot\vec{B}$ operator. Rather, this parameter  $\theta_{\rm eff}$ should be understood as an effective parameter representing the construction of the  $\theta_{\rm eff}$ state for each slice in four dimensional system. In fact, there are three such  $\theta_{\rm eff}$  parameters representing different slices and corresponding external magnetic fluxes. There are similar three $\theta_i$ 
parameters representing the external electric fluxes \cite{Kelnhofer:2012ig}.   This problem of  classification shall not be elaborated in the present work,  as our goal here is to understand and analyze the  simplest possible  topological configurations. We leave the corresponding classification problem which would include  a combination of different BC imposed on  different slices for  future studies. 

Now we are in position to compute the TE for our system in case of non-vanishing external field.
The corresponding generalization of formula (\ref{entropy6}) is given by 
 \be
   \label{entropy_theta}
   && S_{\rm top}(\tau, \theta_{\rm eff})   
   = \ln {\cal{Z}}_{\rm top}(\tau, \theta_{\rm eff}) \nonumber \\
 &-&\frac{1}{2}\chi_{\rm mag}(\tau, \theta_{\rm eff}) 
      +\frac{V \beta}{2}\la B_{\rm ind} (\tau, \theta_{\rm eff}) \ra^2,   
   \ee
   where  $V\equiv L_1L_2L_3$ is 3 volume of the system and  $\langle B_{\rm ind} (\tau, \theta_{\rm eff}) \rangle$  is the induced magnetic field defined as follows \cite{Cao:2013na} 
    \be 
\label{B_ind}
&\,& \langle B_{\rm ind} (\tau, \theta_{\rm eff}) \rangle = -\frac 1 {\beta V}\frac{\partial \ln \mathcal{Z}_{\rm top}(\tau, \theta_{\rm eff}) }{\partial B_{\rm ext}}
 \\
&=& \frac{\sqrt{\tau\pi}}{\mathcal{Z}_{\rm top}}\sum_{k\in\mathbb{Z}}\left(B_{\rm ext}+\frac{2\pi k}{L_1L_2e}\right)\exp{\left[-\tau\pi^2(k+\frac{\theta_{\rm eff}}{2\pi})^2\right]}.\nonumber
\ee
  Magnetic susceptibility $ \chi_{\rm mag}(\tau, \theta_{\rm eff})$ in (\ref{entropy_theta}) is defined similarly to eq. (\ref{entropy6})   with the only difference is that one should  keep  $\theta_{\rm eff}\neq 0$ after taking the  derivatives, i.e.
   \be
   \chi_{\rm mag}(\tau, \theta_{\rm eff})&\equiv& -\frac{2}{\tau}
\frac{\partial^2 \ln {\cal{Z}}_{\rm top}(\tau,\theta_{\rm eff})}{\partial \theta_{\rm eff}^2}.  
      \ee
One can see from (\ref{B_ind}) that   our definition of the induced field  accounts for the total field which includes both terms: the external part as well as the induced topological  portion of the field. In the absence of the external field  when $B_{\rm ext}=0$, the series is antisymmetric under $k\rightarrow -k$ and $\langle B_{\rm ind} (\theta_{\rm eff}=0)\rangle$ vanishes. This feature  is similar to the vanishing expectation value of the topological density (\ref{E2}) in 2d gauge theory  when $\theta=0$. One could anticipate this result   from the very beginning as the theory must respect $\cal{P}$   invariance at $\theta_{\rm eff}=0$, and therefore $\langle B_{\rm ind} \rangle$ must vanish at $\theta_{\rm eff}=0$. 

Now it is easy   compute all the ingredients which enter the expression for entropy (\ref{entropy_theta}) at non-vanishing external field in large $\tau\gg 1$ limit similar to our computations (\ref{entropy_4d}, \ref{chi_4d}). The corresponding asymptotic expressions  for $\theta_{\rm eff}\neq 0$ with exponential accuracy can be represented as follows
\be
\label{asymptotic}
 \ln {\cal{Z}}_{\rm top}(&\tau&\gg 1,\theta_{\rm eff})\rightarrow\frac{1}{2}\ln(\pi\tau)-\pi^2\tau\left(\frac{\theta_{\rm eff}}{2\pi}\right)^2\nonumber\\
  \langle B_{\rm ind} (&\tau&\gg1 ,  \theta_{\rm eff}) \rangle\rightarrow\frac{\theta_{\rm eff}}{eL_1L_2}\nonumber\\
   \chi_{\rm mag}(&\tau&\gg 1, \theta_{\rm eff})\rightarrow 1, 
\ee
where we assume that $|\theta_{\rm eff}|<\pi$. The degenerate case $\theta_{\rm eff}=\pi$ requires a special treatment, similar to 2d analysis presented in section \ref{2d-theta}, and will be discussed at the very end of this section.  
We substitute (\ref{asymptotic}) to general expression for the entropy (\ref{entropy_theta}) to arrive at
\be
   \label{4d_theta}
    S_{\rm top}(\tau\gg 1, \theta_{\rm eff})  \rightarrow \left[\frac{1}{2}\ln({\pi\tau})-\frac{1}{2}\right], 
       \ee
where $-1/2$ in eq. (\ref{4d_theta}) is due to the topologically protected $ \chi_{\rm mag}$ similar to previous  formula (\ref{entropy_4d}) derived for  vanishing background field, $\theta_{\rm eff}=0$. 

The expression (\ref{4d_theta}) for TE   at asymptotically large $\tau$ is independent on $\theta_{\rm eff}$, similar to our  previous   studies in two dimensional ``empty" gauge theory in section \ref{2d-theta}. 
 It implies that   the   topologically protected contribution in eq. (\ref{4d_theta})  assumes exactly the same value (\ref{e})   independently of a magnitude of the external field. Therefore,  we interpret $-1/2$ in eq. (\ref{4d_theta}) as topological  entropy  similar to our discussions leading to  eq (\ref{e}). We claim that the relation (\ref{e}) holds even in the presence of external field  when  $\theta_{\rm eff}\neq 0$. 

\subsection{Degeneracy at $\theta_{\rm eff}=\pi$}\label{doubling} 
In this subsection we want to  analyze a special, but important case with $\theta_{\rm eff}=\pi$ when the system becomes degenerate.
This case is very similar to  our previously studied   system   of  two dimensional ``empty" theory discussed  at the end of  section \ref{2d-theta}.  The crucial element is that our system   is $2\pi$ periodic  as explicit expression  for the partition function (\ref{Z_dual1}) shows. At the same time the point $\theta_{\rm eff}=\pi$ requires a  special treatment as the system shows two-fold degeneracy at this point. Indeed, 
the partition function in vicinity $\theta_{\rm eff}\simeq \pi\pm \epsilon$ at large $\tau$ 
can be approximated  as follows
\be 
\label{Z_pi}
 && {\cal{Z}}_{\rm top}(\tau\gg 1,  \theta_{\rm eff}=\pi\pm\epsilon)\\
 &=&\sqrt{\pi\tau}  \left[e^{ -\pi^2\tau \left(\frac{\pi\pm\epsilon}{2\pi}\right)^2}+  e^{ -\pi^2\tau \left(1-\frac{\pi\pm\epsilon}{2\pi}\right)^2}\right], \nonumber
\ee
where we keep only two leading terms at large $\tau$. One can explicitly see that these two terms identically coincide when 
$\theta_{\rm eff}= \pi $ which implies the degeneracy of the system. These two degenerate states are classified by different directions of the induced magnetic  field characterizing the system. Indeed, with exponential accuracy one gets
\be
\label{B_pi}
 \langle B_{\rm ind} (\tau\gg 1,  \theta_{\rm eff}=\pi-\epsilon) \rangle&=&\frac{\pi}{eL_1L_2} \\
 \langle B_{\rm ind} (\tau\gg 1,  \theta_{\rm eff}=\pi+\epsilon) \rangle&=& -\frac{\pi}{eL_1L_2},  \nonumber
\ee
where two terms in (\ref{B_pi}) are originated   from different  terms in eq. (\ref{Z_pi}) by approaching $\theta_{\rm eff}=\pi $ from different sides. The effect  of degeneracy is very similar in spirit to our studies of  two dimensional ``empty" theory   at the end of  section \ref{2d-theta}. Using an explicit expression for the partition function (\ref{Z_pi}) in vicinity $\theta_{\rm eff}\simeq \pi$ at large $\tau$  one can  arrive to the following additional term  
to the entropy (\ref{4d_theta}) at $ \theta_{\rm eff}=\pi$, 
\be
 \label{ln2}
 \Delta S_{\rm top}(\tau\gg 1,  \theta_{\rm eff}=\pi) =\ln 2.
 \ee
 Few comments are in order. First, we should emphasize that the fact of  degeneracy itself   does not actually depend on magnitude of $\tau$,  though the  expression  (\ref{ln2}) is computed in the limit of  large $\tau\gg 1$.  Indeed, from initial formula (\ref{Z_eff}) one can explicitly see that  for $\theta_{\rm eff}= \pi $ an each term with given positive $k$ has its partner   $(-k-1)$ which produces an  identical contribution to ${\cal{Z}}_{\rm top}(\tau, \theta_{\rm eff}=\pi)$. This   obviously implies the emergence of degeneracy in  the system at $\theta_{\rm eff}= \pi $ which is reflected by (\ref{B_pi}) and (\ref{ln2}). Second,  in the  limit $L_1L_2\rightarrow\infty$ in eq. (\ref{B_pi}) the expectation values of the local operator in these degenerate states are the same (they   vanish,  $\langle B_{\rm ind}\rangle=0$). A proper interpretation in this limit should be formulated in terms of total flux determined by the global behaviour, rather than by local expectation values (\ref{B_pi}):
 \be
\label{theta_pi}
\frac{eL_1L_2}{2\pi} \langle B_{\rm ind}  \rangle_+&=&  \langle \frac{e}{2\pi}\oint A_i dx_i\rangle_{\theta_{\rm eff}=\pi-\epsilon}  =+\frac{1}{2}\\
  \frac{eL_1L_2}{2\pi} \langle B_{\rm ind}   \rangle_-&=&   \langle \frac{e}{2\pi}\oint A_i dx_i\rangle_{\theta_{\rm eff}=\pi+\epsilon}=-\frac{1}2.  \nonumber
\ee
 In other words, these degenerate states can not be distinguished locally; they are classified by the global chracteristics (\ref{theta_pi}), similar to topologically ordered CM systems  \cite{Wen:1989iv,Wen:1990zza,Moore:1991ks,BF,Cho:2010rk,Wen:2012hm,Sachdev:2012dq}. Third, the entropy (\ref{ln2}) does not vanish in thermodynamical limit in the presence of the external magnetic field at $\theta_{\rm eff}=\pi$ as a reflection of the degeneracy of the ground state (\ref{theta_pi}). This  feature is a direct analog of the degeneracy discussed in ref.  \cite{Kelnhofer:2012ig} in the presence of the electric fluxes. As we already mentioned after eq. (\ref{Z_dual1}) one should expect six   different $\theta$'s parameters corresponding 
 three different magnetic fluxes and three different electric fluxes. There will be extra degeneracy (and extra $\ln 2$ contribution to the entropy) when each flux assumes $1/2$ of its value.

Finally, a similar  degeneracy   does not occur for another   $\cal{CP}$ even state with $\theta_{\rm eff}=0$. Indeed, while each contribution with positive $k\neq 0$ has its partner with negative $-k$ in eq. (\ref{Z_eff}) there is a unique   single term  with $k=0$ which  does not have its  partner. This single term   prevents   the   degeneracy to occur in the system with $\theta_{\rm eff}=0 $.

 \section{``BF" formulation of the Maxwell system}\label{BF}
 In the previous section we presented a number of arguments suggesting that the Maxwell system defined on a compact manifold behaves very much in the same way as a topologically ordered system. The arguments include  the analysis of  such ``signatures" of a topological phase  as the degeneracy and the topologically protected finite correction to the entropy. Still, it would be highly desirable to describe the same  system in more conventional way in terms of auxiliary fields governed by  the topological  Chern -Simons action. In this case our claim (that  the Maxwell system defined on a compact manifold is a topologically ordered system)  would be less puzzling and mysterious notion. We should remark here that our system supports conventional massless photons with physical polarizations, in apparent contrast with conventional description of topologically ordered  systems  which normally are characterized by a gap. However, as we already mentioned  the massless photons with physical  transverse polarizations completely decouple from our topological fluctuations described by the topological portion of the partition function ${\cal{Z}}_{\rm top} $ according to eq. (\ref{decouple}).  Therefore, these massless   photons can be completely ignored in our studies  of the topological properties of the partition function, similar to  decoupling of the massless phonons (which are  in fact  responsible for the mere  existence  of a  gap)  in treatment of the topologically ordered superconductors \cite{BF}.  
 
 \subsection{Partition function in ``BF" formulation}
 We wish to derive the topological action for the Maxwell system by using the same conventional  technique exploited e.g.  in  \cite{Zhitnitsky:2013hs} for the so-called ``deformed QCD" and in  \cite{BF} for the Higgs model. Our starting point is to  insert  the  delta function  into the path  integral with the field $b^z(\mathbf{x})$ acting as a Lagrange multiplier
 \be
 \label{delta}
 &&\delta \left[B^z(\mathbf{x})-\epsilon^{zjk} \partial_{j} a_k(\mathbf{x})\right]\sim \\
&& \int {\cal{D}}[b_z] e^{ iL_3\beta\int d^2\mathbf{x} ~b_z(\mathbf{x})\cdot  \left[ B^z(\mathbf{x})-\epsilon^{zjk} \partial_{j} a_k(\mathbf{x})\right]} \nonumber
  \ee
 where $B^z(\mathbf{x})$   in this formula is treated as the original expression for the field operator entering the action (\ref{action4d}),  
		including all  classical k-instanton configurations (\ref{topB4d},\ref{action4d2}) and quantum fluctuations surrounding these classical configurations. In other words, we treat  $B^z(\mathbf{x})$ as fast degrees of freedom. At the same time  $a_k(\mathbf{x})$  is treated as a slow-varying   external source effectively describing the large distance physics  for a given instanton configuration.  Our task now is to integrate out the original   fast ``instantons"  and describe the large distance physics in terms of slow varying fields $b_z(\mathbf{x}), a_k(\mathbf{x})$ in form of the effective action. We use the same procedure by summation over k-instantons 
as   before which is expressed in terms of the  partition function (\ref{Z4d}).  The only new element 
in comparison with the previous computations is that  the fast degrees of freedom must be integrated out in the presence of the new slow varying  background fields   $b_z(\mathbf{x}), a_k(\mathbf{x})$ which appear in  eq. (\ref{delta}). Fortunately, the computations can be easily performed if one notices that the background field $b_z(\mathbf{x})$ enters eq. (\ref{delta}) exactly in the same manner as external magnetic field   enters  (\ref{Z_eff}).
Therefore, assuming that $b_z(\mathbf{x}), a_k(\mathbf{x})$ are slow varying background fields we arrive to the following 
  expression for the   partition function:
  \be
\label{Z_BF}
{\cal{Z}}_{\rm top}  =\sqrt{\pi\tau} \int {\cal{D}}[b_z]  {\cal{D}}[a]  e^{-\pi^2\tau \cdot \int \frac{d^2 \mathbf{x}}{L_1L_2}
\left(\frac{\phi (\mathbf{x})}{2\pi}\right)^2-S_{\rm top}},~~
\ee
where $\phi (\mathbf{x})\equiv eL_1L_2 b_z (\mathbf{x})$  represents the slow varying background auxiliary $b_z$ field
which is assumed to lie in the lowest $k=0$ branch,    $|\phi (\mathbf{x})|< \pi $.  Correspondingly, 
in formula (\ref{Z_BF})   we kept only asymptotically leading term    with $k=0$ in  the series (\ref{Z_eff}) at   large $\tau\gg 1$.  
The topological term  $S_{\rm top}[b_z, a_k] $ in eq. (\ref{Z_BF}) reads  
\be
\label{S_top}
 S_{\rm top}[b_z, a_k] 
 =  i L_3\beta\int d^2\mathbf{x} \left[b_z(\mathbf{x}) \epsilon^{zjk} \partial_{j} a_k(\mathbf{x})\right].~~~~~
\ee
Our observation here is as follows. The topological term (\ref{S_top}) which emerges as an effective description of our system 
is in fact a   Chern-Simons like  topological action. In our simplified setting we limited ourself by considering the fluxes-instantons along $z$ direction only. It is naturally to assume that a more general construction would include fluxes-instantons  in all three directions  which  leads to a generalization of  action   (\ref{S_top}). It is quite natural to expect  that   the action in this case would   assume a  Chern-Simons like form $i \beta\int d^3\mathbf{x} \left[\epsilon^{ijk} b_i(\mathbf{x}) \partial_{j} a_k(\mathbf{x})\right]$ which replaces (\ref{S_top}).
A similar structure in CM systems is known to describe a  topologically ordered phase.  	Therefore, it is not really a surprise that we found in section \ref{4d} some signatures of the topological phases (such as degeneracy and topological entropy) in the Maxwell system defined  on a compact manifold. The emergence of the topological Chern-Simons action (\ref{S_top})  further supports our basic claim that the Maxwell system  on a compact manifold belongs to  a topologically ordered phase. 

 \subsection{Magnetic susceptibility in ``BF" formulation}\label{BF}
 Our goal here is to consider a simplest  application of the effective low energy topological action constructed above (\ref{Z_BF}). To be more specific,  
  we   want to reproduce our expression for  the   magnetic  susceptibility (\ref{chi_mag},\ref{chi_4d}) by integrating out the $b_z$ and $a_k$ fields  using low energy effective description  (\ref{Z_BF}):
  \be
  \label{magnetic1}
  \  \la B_z(x), B_z(0)\ra =\frac{1}{\cal{Z}}  \int {\cal{D}}[b_z]  {\cal{D}}[a]  e^{-S_{\rm tot}[b_z, a_k] } \nonumber\\
  \cdot \left[ \epsilon^{zjk} \partial_{j} a_k(\mathbf{x}) ,\epsilon^{zj'k'} \partial_{j'} a_{k'}(\mathbf{0})\right], 
  \ee
 where $S_{\rm tot}[b_z, a_k] $ determines the dynamics of  auxiliary $b_z$ and  $a_k$ fields, and  it is given by
 \be
 \label{S_tot}
S_{\rm tot}= L_3\beta\int d^2\mathbf{x}\left[ \frac{1}{2} b_z^2(\mathbf{x})
 +i b_z(\mathbf{x}) \epsilon^{zjk} \partial_{j} a_k(\mathbf{x})\right].~
 \ee
 The obtained   Gaussian integral (\ref{magnetic1}) over $\int {\cal{D}}[b_z]$ can be explicitly executed, and we are left with the following   integral over $\int {\cal{D}}[a]$
  \be
  \label{magnetic2}
  &&  \la B_z(x), B_z(0)\ra =  \frac{1}{\cal{Z}}  \int  {\cal{D}}[a]e^{- \frac{L_3\beta}{2}\int d^2\mathbf{x} \left[ \epsilon^{zjk} \partial_{j} a_k(\mathbf{x})\right]^2} \nonumber\\
 &\cdot&     \left[ \epsilon^{zjk} \partial_{j} a_k(\mathbf{x}) ,\epsilon^{zj'k'} \partial_{j'} a_{k'}(\mathbf{0})\right].  
  \ee
 The    integral (\ref{magnetic2}) is also gaussian and can be explicitly evaluated  with the following final result
 \be
 \label{final}
  \la B_z(x), B_z(0)\ra= \frac{1}{\beta L_3} \delta^2 (\mathbf{x}). 
 \ee
 Formula (\ref{final})  precisely reproduces our previous  expression (\ref{mag_local}) derived by explicit summation over fluxes-instantons, and without even mentioning any auxiliary topological fields 
 $b_z (\mathbf{x}), a_k (\mathbf{x})$. It obviously demonstrates  a self-consistency of our formal manipulations with auxiliary topological fields. 
 
 Few comments are in order. First of all, the expression (\ref{mag_local}, \ref{final}) for the magnetic susceptibility represents the contact non-dispersive term   which can not be associated with any physical propagating degrees of freedom as it has a ``wrong sign", similar to our discussions for 2d QED (\ref{dispersion}). The nature of this  contact term is very much the same as  in 2d QED (\ref{local}), and it results from  tunnelling transitions between topologically different but physically identical states. As we mentioned in section \ref{top_2} this contact term in 2d QED can be also understood in terms of KS ghost \cite{KS,Zhitnitsky:2011tr}.    
 Secondly, this term is responsible for the topologically protected contribution to the  
 entropy as eq. (\ref{entropy6}) states, and serves as a signal of a topologically ordered phase.  Finally, as we already mentioned in section  \ref{top_2} an analogous  construction also emerges in ``deformed QCD" ~\cite{Zhitnitsky:2013hs} where the auxiliary topological fields, similar in spirit to   $b_z (\mathbf{x}), a_k (\mathbf{x})$ fields from (\ref{S_tot}), 
and which saturate the ``wrong sign" in topological susceptibility  can be identified with the so-called Veneziano ghost.  Our observation here is that in all considered cases the presence of a non-vanishing  contact term in expression for the entropy and some manifestations 
 of a topologically ordered phase are somehow related. A deep understanding  for such a correlation is still lacking.  
  
  \section{Conclusion }\label{conclusion}
  Before we formulate the main results of this work, we want to make few general comments on connection with other related studies.
  \subsection{Connection  with other related studies}
  In this work we discussed a number of very unusual effects in  Maxwell theory formulated on a compact manifold such as 4 torus.   All these  effects are originated from the topological portion of the partition function ${\cal{Z}}_{\rm top}(\tau, \theta_{\rm eff})$
  and   can not be formulated in terms of  conventional  $E\&M$  propagating photons  with two physical polarizations. 
In fact, a strong hint  that something is missing in attempt to describe everything in   terms of the propagating  degrees of freedom (dof) comes from study of the  ``empty" 2d gauge theory discussed in section \ref{topology} when the system can not support any physical propagating dof. Still, all physical effects relevant for this work are already present in 2d  system. The same comment also holds  for 4d Maxwell theory when photons with two physical polarizations are present in the system.  However, their contribution   completely decouple from topological effects studied in  the present work.

The source of these unusual effects  is as follows. When   the Maxwell  system  is quantized  on a compact manifold  one can not completely  remove all unphysical degrees of freedom  from the system as it would result in emergence of the so-called Gribov's ambiguities~\cite{Gribov:1977wm}, see recent paper \cite{Kelnhofer:2012ig} and also some previous relevant discussions \cite{Killingback:1984en,Parthasarathy:1988sa,Kelnhofer:2007zk}.
These ambiguities  
  were   originally discussed for non-abelian gauge theories  in Minkowski space when one tries to completely remove all unphysical degrees of freedom in the Coulomb gauge~\cite{Gribov:1977wm}, but similar   ambiguities also emerge in abelian  Maxwell theory  defined on a nontrivial  manifold  \cite{Killingback:1984en,Parthasarathy:1988sa,Kelnhofer:2007zk}.
  
   In the present work we opted to keep some gauge freedom in our analysis. 
  The corresponding construction is implemented by allowing the boundary conditions to be periodic up to large gauge transformations, 
  which are precisely reflected by the presence of the ``instantons" (\ref{topB4d})   interpolating between topologically different, but physically identical, pure gauge configurations. 
  The same   topological construction has been used previously in four dimensions in study of the topological Casimir effect \cite{Cao:2013na}
  where it has been claimed that there is an  additional contribution to the Casimir force  in Maxwell theory which can not be accounted for by conventional propagating photons with two physical polarizations.  Similar in spirit computations were also  carried out in \cite{Zhitnitsky:2013hs} in weakly coupled ``deformed QCD" where fractionally charged monopoles- instantons describe the tunnelling events between topologically different, but physically identical winding states. 
  In this case one can also argue that the  gapped ``deformed QCD" belongs to a  topologically ordered phase,  related to the ``degeneracy" of these winding states.

 These computations imply  that an extra energy (and entropy),   not associated   with any physical propagating degrees of freedom,  may appear  in some gauge  systems. The extra energy  in all these cases emerges as a result of dynamics of pure gauge configurations at very large distances. This unique feature of the system when extra energy is not related to any physical propagating degrees of freedom was the main  motivation for a proposal   \cite{Zhitnitsky:2011tr,Zhitnitsky:2011aa}  that the observed dark energy in the universe may have  precisely such non-dispersive  nature. 
 Essentially, the  proposal  \cite{Zhitnitsky:2011tr,Zhitnitsky:2011aa}  identifies the observed dark energy with the Casimir type energy, which however is originated not from dynamics of the physical propagating degrees of freedom, but rather, from the dynamics of the topological sectors  which  are always present in gauge systems. 
A de Sitter behaviour of the universe  in this case can be formulated in terms of the auxiliary topological fields which are
similar in spirit to   $b_z (\mathbf{x}), a_k (\mathbf{x})$ fields from (\ref{S_tot})  and which effectively describe the dynamics of the topological sectors in the expanding background \cite{Zhitnitsky:2013pna}.   It would be very exciting if this new type of energy not associated   with    propagating particles could be experimentally measured in a laboratory as suggested in  \cite{Cao:2013na}. Furthermore, one could argue that 
 these  finite contributions (to the entropy and to the energy), not related to any propagating degrees of freedom cannot be removed by any means such as subtraction or redefinition of observables, see Appendix of ref. \cite{Cao:2013na}
 with corresponding arguments. 
    
  \subsection{Main results}
  
  In this work we have discussed  two  novel (topologically protected) contributions to the thermodynamical  entropy.  The analysis of these  additional contributions to the  entropy, not related to the physical propagating photons,   represents   the main result of the present work.  
  There are two types of  extra terms to the entropy which are related to each other as they both originated from  topological portion of the partition function ${\cal{Z}}_{\rm top}(\tau, \theta_{\rm eff})$,  but nevertheless the physical meaning of these contributions are very  different.  
  
  The contribution of the first type   always enters the entropy  with the negative sign. This term is expressed in terms of the topologically protected magnetic susceptibility $\chi_{\rm mag}$,     can  be represented as the infrared-sensitive surface integral,  and emerges even in the absence of external fields as discussed in section  \ref{TEE}.  
Such contributions are   similar in spirit  to  the topological entanglement entropy  in condensed matter systems, which are normally expressed in terms of the quantum dimension ${\cal{D}}$.
  
    The arguments  that such non-dispersive contact contributions to the entropy may emerge in the gauge systems have been discussed  in the literature  long ago, see original ref. \cite{Kabat:1995eq} and few comments and references on  the recent development at the beginning of  section \ref{2d-entanglement}. These contributions to the entropy are obviously very unique.  It is important to emphasize that these terms  do not contradict to any fundamental principles as we discussed in this work. In particular, the total entropy is always positive function, and the ground-state entropy vanishes in the thermodynamical limit (outside of the degeneracy points, see next paragraph). Nonetheless, the presence of these contact non-dispersive terms expressed in terms of the  infrared-sensitive surface integrals signalling that some kind of long range order, not related to propagating degrees of freedom (irrespectively whether they are massless or massive)   may emerge in the system.

  The contribution   to the entropy of the second type is also related to the topological portion of the partition function ${\cal{Z}}_{\rm top}(\tau, \theta_{\rm eff})$. However, these terms  emerge  only in the presence of external   field represented by $\theta_{\rm eff}=\pi$. In this case  this system becomes degenerate as discussed in section \ref{doubling} and, as a result of this degeneracy  the entropy receives an additional topologically protected contribution with the positive sign as eq. (\ref{ln2}) states, see also footnote \ref{e-flux} on generalization of this construction when   the  external electric fluxes along with magnetic   fluxes are included. This degeneracy can be interpreted as a result 
of a spontaneous symmetry breaking of $\cal{P}$  parity when the induced magnetic field may choose one of two possible directions.  The effect can be  observed only globally (\ref{theta_pi}) rather than locally.  

Such behaviour of the system should be contrasted with conventional picture when physical   photons do not directly couple to an  external field, especially when it is rather small, within  a mG range which    corresponds to $L_1\sim L_2\sim 10^{-2}$cm in eq. (\ref{theta_eff}). The coupling with physical photons  occurs only through the fermion loop, which leads to enormous suppression for all effects normally expressed in terms of a non-linear effective Lagrangian. At the same time the coupling of such small  external fields with ``instanton -fluxes" is always of order of unity  even in absence of any charged fermions. Eventually, it results  to the large effects such as emergence of the degeneracy in the system (\ref{ln2}, \ref{theta_pi}). 

Another result of this work is an  explicit demonstration that the topological portion of the partition function 
${\cal{Z}}_{\rm top}(\tau, \theta_{\rm eff})$ can be reformulated in conventional terms using the Chern -Simons like effective description \ref{BF}. We reproduced the contact term in magnetic susceptibility using this effective description. Such a formulation once again supports our main claim that the Maxwell system formulated on 4 torus belongs to a topologically ordered phase. 

It remains to be seen if the system discussed in the present work can be used as a platform for quantum computations similar to the previous well known suggestion \cite{Kitaev:1997wr} as there are many formal similarities between our system and    topologically ordered condensed matter systems, see 
section   \ref{TEE} with some comments on these analogies. The crucial question is, of course, if  one can manipulate with 
topological ``fluxes- instantons" from section \ref{4d}  in the same way as with real quasiparticles in CM systems using the external magnetic field $\sim \theta_{\rm eff}$ from eq. (\ref{theta_eff}), or varying the boundary conditions. The point is that such manipulations may become efficient only for sufficiently small systems as one should deal with a magnitude of a single flux, as numerical estimates for the topological Casimir effect show \cite{Cao:2013na}. 
 Another way to manipulate the  
topological ``fluxes- instantons"  is to couple the topological fields with real propagating quasi-partciles.  This coupling is likely to create the so-called ``edge states", which is a typical manifestation  of topological phases in CM physics. 
We leave all these interesting subjects for the future studies. 

 \section*{Acknowledgements} 
 I am thankful to my former student Max Metlitski for chatting on properties  of  some condensed matter systems
 related to  this work. I am thankful to him for a number of comments and  clarifications which     improve the presentation. 
 This research was supported in part by the Natural Sciences and Engineering Research Council of Canada.


\end{document}